\documentclass{mn2e}

\usepackage[dvips]{graphicx}

\usepackage{amssymb, latexsym, amsopn}
\DeclareMathOperator{\ach}{ar{\,}cosh}
\DeclareMathOperator{\acth}{ar{\,}coth}

\def\beq{\begin{equation}}
\def\eeq{\end{equation}}

\def\bea{\begin{eqnarray}}
\def\eea{\end{eqnarray}}

\def\ben{\begin{enumerate}}
\def\een{\end{enumerate}}

\def\Ol{\Omega_{\Lambda}}
\def\Om{\Omega_{mat}}

\newcommand*{\lt}{Lema\^{\i}tre-Tolman }
\newcommand*{\ltb}{Lema\^{\i}tre-Tolman}
\newcommand*{\h}{$h^{-1}$ Mpc }
\newcommand*{\hb}{$h^{-1}$ Mpc}

\title{Formation of voids in the Universe within the \lt model}

\author[K. Bolejko, A. Krasi\'nski \& C. Hellaby]
{Krzysztof Bolejko$^{1}$\thanks{Current Address: Nicolaus Copernicus Astronomical Center,
Bartycka 18, 00-716 Warszawa,
Poland; E-mail: bolejko@camk.edu.pl},
Andrzej Krasi\'nski$^{2}$\thanks{E-mail: akr@camk.edu.pl} and
Charles Hellaby$^{3}$\thanks{E-mail: cwh@maths.uct.ac.za}
\\
$^1$
Department of Physics, Astronomical Observatory, University of Warsaw,
Aleje Ujazdowskie 4, 00-478 Warsaw,
Poland
\\
$^2$
Nicolaus Copernicus Astronomical Center, Polish Academy of Science,
ul. Bartycka 18, 00-716 Warsaw,
Poland
\\
$^3$
Department of Mathematics and Applied Mathematics,
University of Cape Town,
Rondebosch 7700, Cape Town,
South Africa
}

\pagerange{\pageref{firstpage}--\pageref{lastpage}}
\pubyear{2004}

\begin{document}

\maketitle

\label{firstpage}

\begin{abstract}
 We develop models of void formation starting from a small
initial fluctuation at recombination and growing to a realistic present
day density profile in agreement with observations of voids.  The model
construction is an extension of previously developed algorithms for
finding a Lema\^{\i}tre-Tolman metric that evolves between two profiles
of either density or velocity specified at two times.  Of the 4 profiles
of concern --- those of density and velocity at recombination and at the
present day --- two can be specified and the other two follow from the
derived model.

We find that, in order to reproduce the present-day void density
profiles, the initial velocity profile is more important than the
initial density profile.

Extrapolation of current CMB observations to the scales relevant to proto-voids
is very uncertain. Even so, we find that it is very difficult to make both the
initial density and velocity fluctuation amplitudes small enough, and still
obtain a realistic void by today.
\end{abstract}


\section{Aim}

Voids are vast regions of the Universe with high negative density contrast,
which are fundamental parts of the large scale structure of the Universe.
Although the mean radius of voids is $10$ \hb, they contain only few galaxies.
According to the data from the 2 degree Field Galaxy Redshift Survey
(2dFGRS\footnote{http://www.aao.gov.au/2df}) processed by Hoyle and Vogeley
(2004), about $40 \%$ of the volume of the Universe is taken up by voids.

The aim of this paper is to describe the non-linear growth of voids out of small
initial density and velocity perturbations on a homogeneous background at the
moment of last scattering. We use the inhomogeneous, spherically symmetric dust
(\ltb) model, an exact solution of Einstein's equations.  This paper also
reports on the main factors responsible for the formation of voids, and a
simulation of void evolution is presented.  As follows from our previous papers
(Krasi\'nski and Hellaby 2002, Krasi\'nski and Hellaby 2004), the final state is
sensitive not just to the amplitude, but also to the exact profile of the
initial perturbations.  So although velocity perturbations of relative amplitude
($\Delta V/V$) around $8 \cdot 10^{-3}$ were needed in our models to reproduce
realistic voids, it is still possible that other profiles can be found for which
a smaller initial velocity amplitude will suffice.

\section{A historical overview}

The discovery of large-scale cosmic voids became possible when astronomers
started to measure the distribution of galaxies in space. Since William and John
Herschel's researches, i.e. from XIX century, it was known that galaxies cluster
(for example in the Virgo or Coma cluster). However there could be no certainty
that these clusters were not just caused by the galaxies being projected on the
celestial sphere. This changed with the publication of the Hubble law in 1929.
Five years later Tolman and, immediately after, Sen studied the stability of
the Friedmann models with respect to inhomogeneous perturbations, and concluded
that they are unstable against a rarefaction caused by a negative perturbation
of either the initial density or the initial velocity. Consequently there should
be condensations as well as underdense regions in the Universe.

Nevertheless it was only at the end of 1970s that voids were discovered. But the
very first sign of their existence had been known 20 years earlier. In 1960
Mayall had measured the redshift of 50 galaxies in the Coma cluster (Mayall
1960). His survey covered 33 square degrees of the celestial sphere. The Coma
cluster was also studied by Chincarini and Rood in 1975, but it was Gregory and
Thompson's (1978) large survey that ended with the discovery of a void. Their
survey covered 260 square degrees of the sky and comprised galaxies up to 15th
magnitude, which is equivalent to a radial velocity of approximately 8000 km/s.

Also J\~{o}eveer and Einasto (1978) observed voids in their redshift
survey. At the beginning of 1980s the term 'void' was first used to call
the regions avoided by galaxies (Rood 1988).

Next researchers discovered extremely large regions avoided by galaxies called
supervoids. The classical example is the void discovered by Kirshner et
al. (1981). From the late 1970s, they were observing the galaxies in the Bootes
and Corona Borealis constellation and discovered an almost empty region with the
size of 50-100 \hb, (where $h$ is the Hubble parameter in units of 100
km/s/Mpc)). More up-to-date measurements determine this size to be approximately
60 \h and prove that this region is not entirely empty. Dey et al. (1990)
observed 21 galaxies within this region and estimated the mean density contrast
to be
 \begin{equation}
   \delta = \frac{\rho_{mat} - \overline{\rho_{mat}}}{\overline{\rho_{mat}}}
   \in (-0.84, -0.66)
 \end{equation}
 where $\rho_{mat}$ and $\overline{\rho_{mat}}$ are the density and mean
density of matter in the Universe.

Other discoveries of voids
followed very fast. In 1982, after a five year survey, the CfA (Center
for Astrophysics) galaxy redshift catalogue was finished. Huchra's team
measured redshifts for over 2400 galaxies with luminosity below
magnitude 14.5. The amount of data increased in later surveys. For
example, the current Sloan Digital Sky Survey (SDSS\footnote{www.sdss.org})
covers one-fourth of the celestial sphere and will measure
redshifts for over $10^8$ objects.

\section{Sizes and shapes of the voids}

The size of a void depends on the luminosity of the galaxies that surround it.
The research done by Lindner et al. (1995) has shown that the mean void size, as
estimated by bright elliptic galaxies, varies between 13 and 36 \hb, whereas the
size evaluated by the measurement of fainter galaxies drops to 9 -- 25 \hb. This
phenomenon is known as the void hierarchy, whose characteristics are listed in
Table \ref{tab1} (Lindner et al. 1996).

\begin{table}
 \caption{\label{tab1}
 The void hierarchy (from Lindner et al.1996).
 }
\begin{center}
  \begin{tabular}{ll}
 Type of object & Mean size \\ \hline

 rich clusters (Abell Catalogue) & 100 \h \\
 poor clusters (Zwicky Catalogue) & 37 \h \\
 bright (M $\le$ -20.3) elliptical galaxies & 30 \h \\
 galaxies brighter than M = -20.3 & 23 \h \\
 galaxies brighter than M = -19.7 & 16 \h \\
 galaxies brighter than M = -18.8 & 13 \h \\ \hline
\end{tabular}
\end{center}
\end{table}

According to the data from 2dFGRS, the average radii of voids in NGP (North
Galactic Pole) and in SGP (South Galactic Pole) are 14.89 $\pm$ 2.67 \h
and 15.61 $\pm$ 2.48 \h respectively (Hoyle and Vogeley 2004). The
maximal radius of a sphere inscribed into a void is 12.09 $\pm$ 1.85 \h in
NGP and 12.52 $\pm$ 1.99 \h in SGP. These sizes were estimated using
galaxies whose luminosity $M_{lim} - 5 log(h)$ varied from -18 to -21.

The computer algorithm employed by Hoyle and Vogeley  in their void search used
a similar rule to that invented by El-Ad and Piran (1997). Firstly, the
algorithm decides whether a particular galaxy should be assigned to the wall (a
region of higher density surrounding the void) or to the void itself. Then the
voids are filled with spheres.

It is apparent that the shapes of small voids are close to spherical,
while the largest voids are more irregular. However, they can still be
divided into smaller spherical voids.
Sato (1982 and 1984)
once hypothesized that spherical voids collide with each other as they
expand, to produce Zeldovich's ``pancakes''. The fact that large
nonspherical voids can be divided into spherical regions may thus
indicate that they are conglomerates of smaller spherical voids that had
already collided.
Together with another result, that for voids the spherical shape is
stable (Sato and Maeda 1983),
this shows that the L--T solution is the right device to model voids.

\section{Density contrast}

While considering the above data, the following question arises: are these
regions really empty? Or are some objects just not bright enough to be detected
by galaxy surveys? This issue was investigated by Thun and co-workers in 1987.
They examined the distribution of faint, dwarf and irregular galaxies from the
Nilson UGC catalogue and concluded that faint galaxies do not form any separate
large-scale structures but occur together with bright galaxies. Similar results
were obtained by Peebles (2001), who analyzed the data from the Optical Redshift
Survey (ORS). The conclusion is -- since there are no bright galaxies in voids,
there should be no faint ones, either. Some observational studies of dwarf
galaxies in the nearest neighborhood (Pustil'nik et al. 1995; Popescu et al.
1997; Kuhn et al. 1997; Gorgin and Geller 1999) found galaxies in voids.
However, these were only single objects, so the deduction of Thun and Peebles
was qualitatively right. Today voids are defined as regions of low density, or
as regions avoided by certain type of objects, for example bright galaxies.
According to these definitions, voids do  not have to be empty. Rojas and
co-workers (2003) focused on the photometrical properties of galaxies in the
voids from the SDSS data. They defined the voids as regions of density contrast
$\delta  \le -0.6$. In their study, they measured galaxies up to $z=0.089$. From
155 000 galaxies they extracted a sample of 13742, in which 1010 were galaxies
in voids. Their analysis shows that galaxies in voids are fainter, more blue and
more compact than galaxies in the walls.

\begin{figure}
    \includegraphics[scale=0.4]{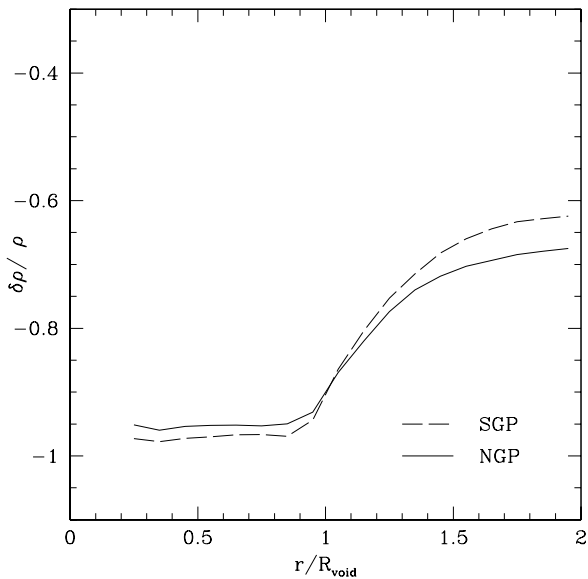}
        \caption{The average density contrast for
        voids in the 2dFGRS, in units of void radius (figure taken from
        Hoyle \& Vogeley 2004).\label{rges}}
\end{figure}

Of the galaxies from the 2dFGRS, 5 $\%$ are void galaxies. The average density
contrast for voids in this survey is presented in Fig \ref{rges}. The mean
density contrast in the central region is $\delta = -0.94 \pm 0.02$ in NGP and
$\delta= -0.93 \pm 0.02$ in SGP.

However, data obtained from observations of galaxies give information about the
luminous matter, while dark matter does not have to concentrate in galaxies.
Peebles (2001), referring to the existing radioastronomical observations of 21
cm waves (sensitive to HI clouds) (Weinberg at al. 1991; Hoffman, Lu and
Salpeter 1992; Szomoru at al. 1996; Zwaan at al. 1997) and to the results of
observations of Lyman $\alpha$ absorption line system (Bergeron and Boisse 1991;
Steidel, Dickinson and Persson 1994; Lanzetta et al. 1995) concluded that not
only galaxies, but also gas clouds avoid voids. This conclusion can only be
justified indirectly because there are no direct observations of Lyman $\alpha$
clouds in voids.

The voids that are discovered in surveys are defined by galaxies with low
redshift. For example, the maximal redshift used by Hoyle and Vogeley (2004) to
find voids in 2dFGRS was $z = 0.138 $. For such a small redshift, the Lyman
$\alpha$ line ($\lambda = 1216$ \r{A}) is in the ultraviolet, which is beyond
the reach of observations from the ground. The resolution of IUE (International
Ultraviolet Explorer) satellite was not sufficient to detect the Lyman $\alpha$
clouds.

One must ask whether luminous matter is a good tracer of mass, or is there a
significant amount of dark matter within voids?  Is the density contrast for
dark matter the same as for luminous matter, or do galaxies prefer regions of
higher density contrast? In the second case, the real density contrast would not
be as low as the one in Fig. \ref{rges}.

Answers to these questions suggested by cold dark matter N-body simulations are
inconclusive. In these simulations, the positions of test particles in a chosen
part of space are traced. Progress in this field depends of the computing power.
In 2002, Arbabi-Bidgoli and M\"{u}ler used $256^3$ particles of masses $10^{11}
{}h^{-1} M_{\odot}$ and $4 \cdot 10^{10} {} h^{-1} M_{\odot}.$ They obtained
density contrasts from $\delta \sim -0.6$ for the voids with the mean diameter
10 - 16 \h to $\delta \sim -0.8$ for the voids with mean diameter 36 - 50 \h.
Benson with co-workers (2003) used $512^3$ particles of masses $10^9 ~h^{-1}
M_{\odot}$. Their simulations implied that the density contrast should be
$\delta \sim -0.8, 0.85$. The Gottl\"{o}ber team (2003) used $1024^3$ particles
of $4 \cdot 10^7 ~h^{-1} M_{\odot}$ and obtained the density contrast of
approximately $\delta = - 0.9$.  Some N-body simulations face a crucial problem,
as they predict that voids should be filled with dwarf galaxies.

In our models, we assume that the real density contrast in voids is the same as
the one obtained from galaxy observations.

\section{The Lema\^itre -- Tolman model}

\subsection{Formulae and general properties}

The \lt model is a spherically symmetric solution of Einstein's equations with a
dust source. In comoving and synchronous coordinates, the metric is:
\begin{equation}
{\rm d}s^2 =  c^2{\rm d}t^2 - \frac{R_{,r}^2(r,t)}{1 + 2 E(r)}\ {\rm
d}r^2 - R^2(t,r) {\rm d} \Omega^2, \label{ds2}
\end{equation}
where $ {\rm d} \Omega^2 = {\rm d}\theta^2 + \sin^2 \theta {\rm d}\phi^2$,
and $E(r)$ is an arbitrary function of $r$. Because of the signature
$(+, -, -, -)$, this function must obey $E(r) \ge - \frac{1}{2}.$

The Einstein  equations reduce to the following two:
\begin{equation}\label{LTden}
\kappa \rho c^2 = \frac{2M_{,r}}{R^2 R_{,r}},
\end{equation}
\begin{equation}\label{LTevol}
\frac{1}{c^2}R_{,t}^2 = 2E + \frac{2M(r)}{R} + \frac{1}{3} \Lambda R^2,
\end{equation}
\noindent where M(r) is another arbitrary function and $\kappa = \frac{8 \pi G}{c^4}$.

When $R_{,r} = 0$ and $M_{,r} \ne 0$, the density becomes infinite. This
happens when shell crossings occur. Equation (\ref{LTevol})
 is similar to its Newtonian counterpart
  for a spherical dust distribution:
\[\frac{1}{2}\mathcal{R}_{,t}^2 = \frac{\mathcal{E}}{m} + \frac{G \mathcal{M}}{\mathcal{R}},\]
 where $\mathcal{R}$, $\mathcal{E}$ and $\mathcal{M}$ are respectively
the radial coordinate, the energy of the particles, and the mass within
radius $\mathcal{R}$ (in Newtonian mechanics, the cosmological constant
is not considered). Therefore, $M(r) c^2/G$ is the mass inside the shell
of the radial coordinate $r$, and $E(r) c^2$ is the energy per mass unit.

Equation (\ref{LTevol}) can be solved by simple integration:

\begin{equation}\label{LTsolfort}
\int\limits_0^R\frac{d\tilde{R}}{\sqrt{2E + \frac{2M}{\tilde{R}} + \frac{1}{3}\Lambda
\tilde{R}^2}} = c \left[t- t_B(r)\right], \label{cal}
\end{equation}
where $t_B$ appears as an integration constant, and is an arbitrary function of
$r$. This means that the Big Bang is not a single event as in the Friedmann
models, but occurs at different times for different distances from the origin.

When $\Lambda = 0$, the above equation can be solved in parametric form:

\begin{itemize}

\item
For $E < 0$:
\begin{eqnarray}
R &=&-\frac{M}{2E}(1-\cos\eta), \nonumber \\
\eta - \sin\eta &=& \frac{(-2E)^{3/2}}{M}c(t-t_B(r)).
\label{e<0}
\end{eqnarray}
Eliminating  $\eta$ one can write this as\footnote{Equation (\ref{tben}) applies
with $\eta < \pi$, i.e. in the expansion phase. We do not consider the
recollapse phase in this paper.}
\begin{eqnarray}
ct_B &=& ct - \frac{M}{(-2E)^{3/2}} \left[ \arccos \left(1 + 2\frac{ER}{M}
\right) \right.
 \nonumber \\
    &-& \left. \sqrt{1- \left( 1 + 2\frac{ER}{M} \right)^2} {~}\right]
\label{tben}
\end{eqnarray}

\item
For $E=0$:
\begin{equation}
R=\left[ \frac{9}{2}Mc^2( t-t_B(r))^2 \right]^\frac{1}{3}.
\label{e=0}
\end{equation}

\item
For $E>0$:
\begin{eqnarray}
R &=& \frac{M}{2E}(\cosh\eta - 1), \nonumber \\
\sinh\eta - \eta &=&\frac{(2E)^{3/2}}{M}c(t-t_B(r)).
\label{e>0}
\end{eqnarray}
 or equivalently:
\begin{eqnarray}
ct_B &=& ct - \frac{M}{(2E)^{3/2}} \left[ \sqrt{\left( 1 + 2\frac{ER}{M}
\right)^2 -1} \right.
 \nonumber \\
    &-& \left.
  \textup{arcosh} \left(1 + 2\frac{ER}{M} \right) \right].
 \label{tbep}
\end{eqnarray}

\end{itemize}
Thus, the evolution of a \lt model is determined by three arbitrary
functions: $E(r)$, $M(r)$ and $t_B(r)$. The metric and all the formulae
are covariant under arbitrary coordinate transformations of the form $r
= f(r')$. Using such a transformation, one function can be given a
desired form. Therefore the physical initial data for the evolution of the \lt
model consist of two arbitrary functions.

\subsection{The Friedmann limit}

The \lt model is a generalization of the Friedmann models and becomes a
Friedmann model when the following conditions are fulfilled:
\beq \label{tebeconst}
 \mbox{(i)} \quad t_{B} = \mbox{constant}.
\label{tbc}
\eeq
This constant is usually set to zero.
\beq
 \mbox{(ii)} \quad \frac{|E|^{3/2}}{M} = \mbox{constant}.
\eeq
In the Friedmann limit, the density distribution is a function of the
time coordinate only, and is expressed by the formula:
 \beq
\kappa \rho c^2 = \frac{6M}{R^3}. \label{Weyl}
 \eeq

The above conditions are invariant under any coordinate transformation. In the
class of coordinates used here, one can choose the radial coordinate as:
 \beq
 R(r,t) = r S(t),
\label{r=s}
 \eeq
where $S$ is the scale factor of the Friedmann models, and then
 \beq
 M(r) = M_0 r^3, ~~~~~~~E(r) = E_o r^2.
\eeq

The Friedmann limit is an essential element in our approach. As mentioned above,
our model of void formation describes a single void in an expanding Universe.
Far away from the origin, the density and velocity distributions tend to the
values which they would have in a Friedmann model. The mean sizes of voids
presented in the literature are estimated in the Friedmann models (each estimate
uses one specific model, but for low redshifts the differences between different
models are negligible). Using (\ref{r=s}) one can identify the areal radius of a
void $R(r_v, t_0)$ with the mean void radius given in the literature.

\section{Background models}

The aim of this paper is to describe the formation of voids from initial density
and velocity perturbations at the time of last scattering, and also to check how
the evolution of a void depends on the background model. This requires knowledge
of the density and velocity perturbations, and also of the age of the Universe
at the moment of last scattering. Although not all the background models used
here are consistent with observations, even those excluded by observations will
clarify some mechanisms responsible for void formation. The astronomical data
put limits on the values of some parameters. From the observation of the oldest
stars the lower limit for the age of the Universe is estimated to be
approximately $12 - 14 \times 10^9$ years (Spergel et al. 2003). From the
measurement of the movement of galaxies in clusters and of matter in galaxies
one can estimate the mean matter density. In the critical density units this
value is $\Om \sim 0.3$. Observations of type Ia supernovae and of the microwave
background radiation suggest a nonzero cosmological constant, of approximate
value $\Ol \sim 0.73$ (Bennett et al. 2003).

In the following, the subscript ${}_b$ indicates a background value, ${}_0$
indicates a present day value, and ${}_{ls}$ indicates a value at last
scattering.

\subsection{Background models without cosmological constant }

\subsubsection{The Einstein-de Sitter Universe}

The Einstein-de Sitter universe is the flat Friedmann model filled with matter
and without the cosmological constant. This model can be obtained from the \lt
model. From (\ref{Weyl}) and (\ref{tebeconst}) with $E = 0$ it follows that the
density of the homogeneous background is:
 \beq
\rho_{b} = \frac{6M}{\kappa c^2 R_b^3} = \frac{1}{6 \pi G t^2}. \label{flrw}
 \eeq
Substituting above the critical density obtained from the Hubble constant $H_0 =
72$~km/s/Mpc (Bennett et al. 2003) one can get that the present age of the
Universe is $9.053 \times 10^9$ years. The last scattering photons on electrons
took place when $z \approx 1089$ (Bennett et al. 2003). It was not a single
event, but a process extended in time. In what follows, it will be assumed that
the last scattering took place when $z=1089$. At that instant the background
density was equal to
 \beq
(\rho_{ls})_b = (\rho_0)_b (1 + z)^3 \label{rhobeg}
 \eeq
so from (\ref{flrw}) we get that the Universe was 252 000 years old.

The following quantity is a measure of the velocity:
 \beq
 b= \frac{R_{,t}}{c M^{1/3}}.
 \eeq
In the flat background this becomes:
 \beq
b_b = \frac{(R_{,t})_b}{c M^{1/3}} = \left(\frac{4}{3 ct} \right)^{\frac{1}{3}}.
 \eeq

\subsubsection{The hyperbolic background }

The hyperbolic background is the $k < 0$ Friedmann model. The age of the
Universe at the moment of last scattering can be calculated using eqs.
(\ref{e>0}). In the Friedmann models the factor $M/E^{3/2}$ is constant and one
can calculate it assuming that the current expansion rate of the homogeneous
background is given by the Hubble constant:
 \beq
  H_0^2 = \left. \frac{R_{,t}{}^2}{R^2} \right|_0 = \frac{2Ec^2}{R_0^2} +
\frac{2Mc^2}{R_0^3} = \frac{2Ec^2}{R_0^2} + \frac{1}{3} \kappa c^4
\rho_0,
 \eeq
From this one gets:
\begin{equation}
\frac{2E}{M^{2/3}} = \frac{1}{3} \kappa c^2 ( \rho_{crit} - \rho_0)
\left( \frac{6}{ \kappa \rho_0 c^2} \right) ^{2/3},
 \label{epm}
\end{equation}
where $\rho_{crit} = 3 H_0^2 / \kappa c^4$ is the density of a $k = 0$ Friedmann
model. Substituting the above result in the first of (\ref{e>0}), one gets
 \beq
 \eta_{ls} = \ach \left( 2 \frac{1+ \Omega_{mat}}{\Omega_{mat} (1+z)} \right),
 \eeq
 \noindent
where $\Omega_{mat} = \rho_0/\rho_{crit} = \kappa c^4 \rho_0 / 3 H^2$. From the
above formula, and from (\ref{epm}) and (\ref{e>0}) it follows that the age of
the Universe with $\Om = 0.27$ when $z = 1089$ is equal to 477 000 years. The
present age is $11.1 \times 10^9$ years. In the model with $\Om = 0.391$ (this
model has a similar present day velocity
 to the model with $\Om = 0.27$ and $\Ol = 0.73$) one can obtain 402 000
years for the instant of last scattering and $10.6 \times 10^9$ years
for today. The value of the background density was calculated from eq.
(\ref{rhobeg}).

The background velocity in the hyperbolic model  is:
 \bea
 b_b & = & \frac{(R_{,t})_b}{c M^{1/3}}
 = \sqrt{\frac{2E}{M^{2/3}} + \frac{2M^{1/3}}{R_b}} \nonumber \\
 & = & \sqrt{ \frac{2E}{M^{2/3}}
 + 2 \left(\frac{1}{6} \kappa \rho_b c^2 \right)^{1/3}}.
 \eea

\subsection{The elliptic background}

The elliptic background is the $k > 0$ Friedmann model. The procedure is
similar to the one above. To calculate the age of the Universe one must
use eqs. (\ref{e<0}),
  and the value of $\eta$ is
\[ \eta_{ls} = \arccos \left( 1- 2 \frac{\Omega_{mat} -1 }{\Omega_{mat} (1+z)}
\right).\]
The age of the Universe in the model with $\Om =11$ is 76 000 years at
last scattering and $4.6 \times 10^9$ years today.

\subsection{Background models with the cosmological constant }

\subsubsection{The flat background }

When $E=0$, eq. (\ref{LTsolfort}) can be solved explicitly:
 \bea
 \int \limits_{0}^{R} dR' \frac{1}{ \sqrt{ \frac{2M}{R} + \frac{1}{3} \Lambda R^2}} \nonumber \\
 = \sqrt{\frac{4}{3 \Lambda}} \acth \left( \frac{3}{ \Lambda} \sqrt{\frac{2M}{R^3} + \frac{ \Lambda}{3}} \right)
 = c(t - t_B).
 \eea
 or equivalently:
 \beq R^3 = \frac{6M}{ \Lambda} \sinh^2
\left(ct \sqrt{\frac{3 \Lambda}{4}} \right).
 \eeq
It follows that the background density is:
\begin{equation}
 \kappa c^2 \rho_b = \frac{\Lambda}{ \sinh^2 \left(ct \sqrt{\frac{3 \Lambda}{4}} \right)}.
\label{gtla}
\end{equation}
The age of the Universe can be calculated from the formula given above.
Substituting the values suggested by astronomical observations, $\Om =
0.27$ and $\Omega_{\Lambda} = \frac{\Lambda c^2}{3 H_0^2} = 0.73$ we
obtain that the density is equal to the critical density, the Universe
is $13.48 \times 10^9$ years old today, while at the moment of last
scattering the age of the Universe was 484 000 years.

The background velocity in the flat Universe with the cosmological
constant is given by the formula:
 \beq
 b_b = \frac{(R_{,t})_b}{c M^{\frac{1}{3}}}
 = \frac{2}{3} \left( \frac{6}{\kappa c^2  \rho_b} \right)^{\frac{1}{3}}
 \sqrt{ \frac{3 \Lambda}{4}} \coth^2 \left(ct \sqrt{\frac{3 \Lambda}{4}} \right)
 \eeq

\subsubsection{Non-flat background models}\label{nonflat}

We follow an analogous procedure to the $\Lambda = 0$ case, and calculate:
 \beq \frac{2E}{M^{2/3}} = \frac{1}{3} \kappa \rho_{crit} c^2 ( 1 -
\Omega_{mat} - \Omega_{\Lambda}) \left( \frac{6}{ \kappa \rho_0 c^2}
\right) ^{2/3},
 \label{epmzl}
 \eeq
where $\Omega_\Lambda = (\Lambda / \kappa c^2) / \rho_{crit} = \Lambda c^2 / 3
H^2$. When $E \ne 0$, eq. (\ref{LTsolfort}) does not have an analytic solution.
Let us denote:
\[ \frac{R_b}{M^{1/3}} = a,~~~~~~~~~~~\frac{2E}{M^{2/3}} = \alpha,\]
then eq. (\ref{LTsolfort}) can be written:
 \beq
\int\limits_0^a \frac{dx}{\sqrt{\alpha + \frac{2}{x} +
\frac{1}{3}\Lambda x^2}} = ct.
\label{age}
 \eeq

In the model with $\Om = 0.27$ and $ \Ol = 1.64$ one finds that the age
of the Universe at the moment of decoupling was 485 000 years and at
present it is $32.46 \times 10^9$ years. The background velocity is:
\begin{eqnarray}
  b_b &=& \frac{(R_{,t})_b}{c M^{1/3}} = \sqrt{
\frac{2E}{M^{2/3}} + \frac{2M^{1/3}}{R_b} + \frac{1}{3}
\Lambda \frac{R_b^2}{M^{2/3}}}
\nonumber \\
&=& \sqrt{ \frac{2E}{M^{2/3}}
+ 2 \left(\frac{1}{6} \kappa \rho_b c^2 \right)^{1/3}
+ \frac{1}{3} \Lambda \left( \frac{6}{ \kappa \rho_b c^2} \right)^{2/3}}.
 \label{velbeg}
\end{eqnarray}

  The parameters of the various background models are summarized in Table
 \ref{Age}.

\begin{table}
 \caption{\label{Age}
The age of the Universe at the present epoch and at the moment of last
scattering
 }
\begin{center}
  \begin{tabular}{lll}
 Model & Present age of & Age at last \\
 & the Universe [y] & scattering [y]  \\ \hline

 $\Om = 1 $, $\Ol = 0$ & $9.053 \times 10^{9} $ & $252 \times 10^3 $ \\
 $\Om = 0.27 $, $\Ol = 0$ & $11.1 \times 10^{9}$  & $477 \times 10^3$  \\
 $\Om = 0.391 $, $\Ol = 0$ & $10.6 \times 10^{9}$  & $402 \times 10^3$  \\
 $\Om = 11 $, $\Ol = 0$ & $4.6 \times 10^{9}$  & $76 \times 10^3 $ \\
 $\Om = 0.27 $, $\Ol = 0.73$ & $13.48 \times 10^{9}$  & $484 \times 10^3$  \\
 $\Om = 0.27 $, $\Ol = 1.64$ & $32.46 \times 10^{9} $ & $485 \times 10^3 $ \\
  \hline
\end{tabular}
\end{center}
\end{table}

\section{Perturbations at the moment of last scattering}

The microwave background radiation is a relic from the epoch when the Universe
was young and hot. When the temperature of the Universe dropped below 3000 K,
the free path of photons became comparable to the Hubble radius and radiation
stopped interacting with matter. Consequently, an analysis of this radiation can
give us information about the state of matter at that moment. The spectrum of
the cosmic microwave background is the spectrum of the black body with the mean
temperature of $T \approx 2.725$ K (Mather et al. 1999). More precise
observations measure the fluctuations with amplitude $ \delta T/T \sim 10^{-5}
$. These are related to the density fluctuations at last scattering.
Unfortunately, this relation is complicated -- one must take into account
several effects.

\subsection{Linear and angular diameters }
\label{ladiam}

\begin{figure}
    \includegraphics[scale=0.35]{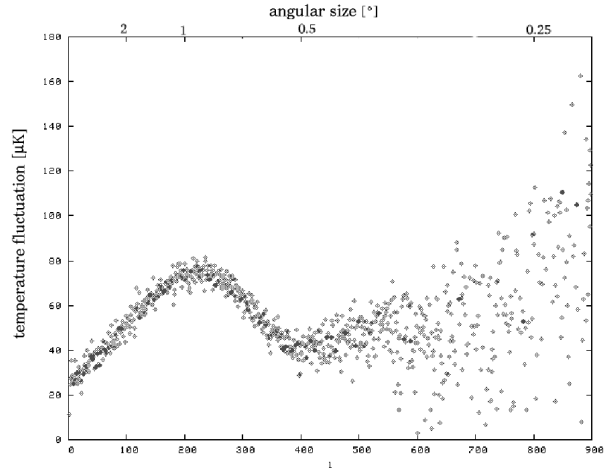}
    \caption{
  The temperature fluctuations ($\Delta T$)
    measured by WMAP (based on
    data from http://lambda.gsfc.nasa.gov/product/map/).\label{wmap}}
\end{figure}

The observations of the microwave background radiation measure differences in
temperature between two points of the celestial sphere. The angular resolution
of the instruments is not high. The Wilkinson Microwave Anisotropy Probe
satellite (WMAP) measures the radiation in five different ranges and, depending
on the range, has the resolution from $0.82^{\circ}$ to $0.21^{\circ}$ (Bennett
et al. 2003). The results of the temperature measurement are presented in Fig.
\ref{wmap}.

The following question arises: what are the linear diameters of the
regions that WMAP can still see as separate? And is this resolution
sufficient to observe the regions which evolved into the currently
observed voids?

The linear diameters can be estimated using the scale law of the
Friedmann models:
 \beq
\frac{L_0}{L_{dec}} = \frac{ \chi (r) S_0}{ \chi (r) S_{dec}} = 1 + z.
 \eeq
where $\chi (r)$ is the distance in the space $t =$ const:
\beq
\chi(r) := (1/S(t)) \int\limits_0^r g_{rr}(t, r') {\rm d} r'.
 \eeq
These diameters are related to the angular diameters:
 \beq
L = D \Delta \theta,
 \eeq
where $D$ is the angular distance; in the Friedmann models it is determined by:
 \beq
D= \frac{S_o}{1+z} \mathcal{F}(d),
 \eeq
where
\begin{equation}
\mathcal{F}(d) = \left\{
\begin{array}{ccc}
\sin (d), & $when$ & \Om + \Ol > 1 \\
d & $when$ & \Om + \Ol = 1 \\
\sinh (d) & $when$ & \Om + \Ol < 1
    \end{array} \right.
\end{equation}
and $d$ is given by the formula:
\beq
d = \frac{c}{H_0 S_o}
\int\limits_0^z dz'  \frac{1}{\sqrt{\Omega_c(1+z')^2 + \Omega_{mat}(1+z')^3 +
\Omega_{\Lambda}}},
\eeq
where $ \Omega_c = 1 - \Omega_{mat} - \Omega_{\Lambda}$.
Assuming that the moment of last scattering was when $z = 1089$, and
that the mean void diameter is $25$ \hb, for different models one gets the
results shown in table \ref{tab2}.

\begin{table}
\caption{Angular diameters of the pre-void region}
\label{tab2}
\begin{center}
 \begin{tabular}{@{}lc}
 Model & $\Delta \theta$ \\ \hline

 $\Om=1,~~ \Ol =0$  & $0.246^{\circ}$  \\
 $\Om=0.27,~~ \Ol =0$ & $0.071^{\circ}$ \\
 $\Om=0.27,~~ \Ol = 0.73$  & $0.143^{\circ}$ \\ \hline
\end{tabular}
\end{center}
\end{table}

Comparing data from Table \ref{tab2} with Fig. \ref{wmap} one sees that
the resolution of the WMAP instruments is not sufficient to make direct
measurements of the temperature fluctuations in the regions that became
voids.

\subsection{Initial fluctuations }\label{infl}

As the starting point of further considerations, we need a rough estimate
of the initial conditions.
The procedure for estimating the density fluctuations from the observed
temperature fluctuations is complicated. The main contributions to the
observed temperature perturbation are from the intrinsic temperature
fluctuations of the emitting fluid, the frequency shift as light emerges
from a fluctuation in the gravitational potential (Sachs-Wolfe effect),
and from the Doppler effect (motion of the emitter).
The moment of last scattering also depends on the cosmological
background model, but for different models the differences are
negligible.
(For completeness, one should take into account what happened to the
radiation on the way to the observer: the integrated Sachs-Wolfe effect ---
the effect of gravitational perturbations along the line of sight, the
Rees-Sciama effect --- the influence of changes of the gravitational
potential with time, the Sunyaev-Zeldovich effect --- radiation
interacting with galaxy clusters, and also the non-linear
consequences of the photons travelling in an inhomogeneous space.)

It is found (e.g Padmanabhan 1996)
that the intrinsic temperature fluctuations
obey $(\Delta T/T)_{intrinsic} \sim \Delta \rho / 3 \rho$.  The magnitude of
velocity perturbations is usually not given%
 \footnote{
 It is commonly stated by specialists that $\Delta V / c \sim 10^{-5}$ but
we have not been guided to a reference that says this or something equivalent.}%
 .
Fluctuations in the fluid motion of magnitude $\Delta V$ (away from
uniform expansion) must contribute a Doppler term $(\Delta
T/T)_{Doppler} = \Delta V/c$, so observations of $\Delta T/T \leq
10^{-5}$ put an upper limit of $10^{-5} c$ on $\Delta V$.  On the other
hand, since the fluctuations are acoustic oscillations, the three
contributions to $(\Delta T / T)_{observed}$ must be of comparable
magnitude, $(\Delta T / T)_{intrinsic} \sim \Phi/3 \sim \Delta V / c$.
Indeed for an oscillating fluid with a relativistic equation of state
($\partial p / \partial \rho = c^2$), we must have $\Delta V \sim c
\Delta \rho / 3 \rho$.

However, the major problem is not with calculations, but with the data.
The present data are available for scales larger than the scale of a
single void, and all these calculations must rely on extrapolations.

The WMAP data that was used as the source for Fig. \ref{wmap} has such a large
scatter for scales around $0.2^\circ$ that the temperature fluctuation could be
anything from $10^{-4}$ to $10^{-6}$. Using the extrapolation proposed by the
WMAP team one would get $\Delta T / T \approx 2 \cdot 10^{-5}$.

To estimate $\nu = \Delta V / V_b = \Delta b / b_b$ for a present day scale of
$L_0 = 12$~Mpc, we write $L = L_0/(1 + z)$, $H = H_0 (1 + z)^{1/n}$,
$V_b = L H$ and $n = 2/3$ for an Einstein-de Sitter model.  Thus we find, for
$H_0 = 72$~km/s/Mpc and $z = 1089$, that
 \beq
 \nu \approx  2 \times 10^{-4}.
 \eeq

\section{Formation of voids }

In this section we will check whether it is possible to evolve the voids
from a small initial density and velocity perturbation imposed on a
homogeneous background. We will also check what is the influence on the
structure formation of the following factors:

\ben
\item
The shape and the amplitude of the initial perturbations,
\item
The evolution time,
\item
The expansion rate,
\item
The outflow of mass from central parts of the void,
 \een

The evolution of the void will be calculated with six different background
models. Some of these models, especially the elliptic ones with and without
cosmological constant, are inconsistent with the observations. These are used
not in order to obtain a model of the observed Universe, but to check which of
the factors listed above are more important in the process of void formation.
For comparing the various models, let us assume that the initial conditions are
independent of the background model.

To determine the evolution of the \lt model one needs to know two functions. In
this paper, these functions will be the initial density and velocity
distributions. This is not the only method to specify the evolution in the \lt
model. The evolution can be determined also by giving the initial and the final
density profiles (Krasi\'nski and Hellaby 2002) or the initial velocity
distribution and the final density distribution (Krasi\'nski and Hellaby
2004).\footnote{The numerical examples in those papers used the present day for
the final time.}

\subsection{The algorithm}

The computer algorithm used to calculate the evolution of a void was written in
Fortran and consisted of the following steps.  Numerical methods are from Press
et  al. (1986) and Pang (1997).

\begin{enumerate}

\item   The initial time $t_i$ was chosen to be the time of last scattering.
This moment was calculated as described in section
\ref{nonflat}.  Equation (\ref{age}) was integrated using Bode's rule
with the step of value $10^{-6} \cdot (6 / (\kappa \rho_{cr} c^2 \Om))^{1/3}$.  The homogeneous
background density and velocity were calculated from eqs. (\ref{rhobeg})
and (\ref{velbeg}) as described in section \ref{nonflat}.

 \item   The initial density and velocity fluctuations, imposed on this
homogeneous background, were defined by functions
of radius $\ell$,
 \[   \delta(\ell) ~~~~~~\mbox{and}~~~~~~ \nu(\ell) ,    \]
 as listed in Tables \ref{InitDenPert} and \ref{InitVelPert}, and the actual
density and velocity followed from
 \[   \rho_i(\ell) = (\rho_b)_i ( 1 + \delta(\ell))
      ~~~~~~\mbox{and}~~~~~~
      b(\ell) = (b_b)_i (1 + \nu(\ell)) ,    \]
 with $\rho$ measured in units of $10^{45} M_{\odot}$/kpc$^3$ and $b$ in
kpc$^{-1/3}$. The parameter $\ell$ is defined as the areal radius at the moment
of last scattering, measured in kiloparsecs, and is also used for the radial
coordinate, i.e.
 \[   r = \ell = R_{i}/d = R(r, t_{i})/d   \]
where $d = 1$~kpc.

\item   Then the mass inside the shell of radius $R_i$, measured in
kiloparsecs, was calculated by integrating eq. (\ref{LTden}):
 \begin{equation}
   M(\ell) - M(0) = \left. \frac{\kappa c^2}{2} \int\limits_{\ell_{min}}^{\ell}
   \rho_i(\ell') \ell'^2 d\ell' \right|_{t = t_{ls}} . \label{ModR}
 \end{equation}
Since the density distribution has no singularities or zeros over extended
regions, it was assumed that $\ell_{min} = 0$ and $M = 0$ at $\ell = 0$.

The integration was done using Bode's rule, with step size $2.5$ pc.

\item   $E$ was calculated from $R_i$, $V = (R_{,t})_i$, $M$ and a chosen
$\Lambda$ value, using eq. (\ref{LTevol}).

\item   Then $t_{B}$ was calculated from eq. (\ref{LTsolfort}) using
Simpson integration, with step size $10^{-5} \ell$.

\item   Once $M$, $E$ and $t_B$ are known, the  state of the \lt model can be
calculated for any instant.  Solving eq. (\ref{LTevol}) with the second-order
Runge-Kutta method for $R(t,\ell)$ along each constant $\ell$ worldline, we
calculated the value of $R(t,\ell)$ and $R_{,t}(t,\ell)$ up to the present
epoch.  The time step was $5 \times 10^{5}$ years.

\item   The density $\rho(t,\ell)$ was then found from (\ref{LTden}) using the
five-points differentiating formula. The adjusted differences between adjacent
worldlines, used in estimating derivatives, was 10 pc.

\item   The density contrasts presented in section \ref{voidmod} were estimated
from eq. (\ref{dcon}).

 \end{enumerate}

\begin{table*}
\caption{ \label{InitDenPert} The initial Density Perturbations used in the
runs.  All the values in the table are dimensionless, and the distance parameter
is the areal radius in kiloparsecs $\ell = R_i$/1kpc. Note that the output
figures depend on the initial perturbation in both density and velocity.
  }
\begin{tabular}{llllll}

 Section & Model & Density perturbation & Parameters  & Graph & Output \\ \hline
\ref{inperofhom} & 1-6 &  $  \delta_{INIT}(\ell) = A \cdot ( b \arctan{c}
 - d \ell - $ & $  A = 1.1 \times10^{-5} $ & Fig. \ref{beg} & Fig.
 \ref{konk},\ref{wm} \\
  & & $ f e^{-g^2} - e^{-h^2} - i e^{-j^2}) \cdot k $  &  $ b = 4  $ & &  \\
  & &  &  $ c = 0.16 \ell - 2.2 $ & \\
  & &  &  $ d = \frac{5}{35} $ & \\
  & &  & $ f= 0.5 $ & \\
  & &  & $ g= \frac{\ell-7}{6} $  & \\
  & &  & $ h = \frac{\ell-9}{7} $ & \\
  & &  & $ i= 1.4 $ & \\
  & &  & $ j= \frac{\ell - 11}{3} $ & \\
  & &  & $ k= \frac{1}{1 + 0.03 \ell} $ & \\
  \hline
\ref{homovel} & 1-6 &  $  \delta = \delta_{INIT}(\ell) $ & & Fig. \ref{beg} &
Fig. \ref{kmbp} \\
 \hline
\ref{homorho} & 1-6 &  $  \delta = 0 $ &  &  & Similar to Figs.
\ref{konk},\ref{wm} \\
 \hline
\ref{amplitude} & 1-6 & $ \delta_{AMP}(\ell) = A \cdot ( b \arctan{c}
 - d \ell - $ & $  A = 7.5 \times10^{-4} $ & & Fig. \ref{kmbc} \\
  & & $ f e^{-g^2} - e^{-h^2} - e^{-j^2}) \cdot k $  &  $ b = 4  $ &  \\
  & &  &  $ c = 0.08 \ell - 1.1 $ & \\
  & &  &  $ d = \frac{5}{55} $ & \\
  & &  & $ f= 0.4 $ & \\
  & &  & $ g= \frac{\ell}{4} $  & \\
  & &  & $ h = \frac{\ell-2}{7} $ & \\
  & &  & $ j= \frac{\ell - 4}{3} $ & \\
  & &  & $ k= \frac{1}{1 + 0.03 \ell} $ & \\
 \hline
\ref{modelfit} & 1,2,3 &  $  \delta_{1,2,3}(\ell) = 100 \cdot
 \delta_{INIT}(\ell) $ &  & Fig. \ref{plb} & Fig. \ref{glb},\ref{hlb},\ref{fit}
 \\
 \hline
\ref{modelfit} & 4 &  $  \delta_{4}(\ell) = 2 \cdot
 \delta_{1,2,3}(\ell) $ & & Fig. \ref{plb} & Fig. \ref{glb},\ref{hlb},\ref{fit}
 \\
 \hline
\ref{radtt} & both &  $  \delta_{RAD}(\ell) = A \cdot ( b \arctan{c}
 - d \ell - $ & $  A = 7.5 \times10^{-4} $  &  & Fig. \ref{grad} \\
  & & $ f e^{-g^2} - e^{-h^2} - e^{-j^2}) \cdot k $  &  $ b = 4  $ &  \\
  & &  &  $ c = 0.08 \ell - 1.1 $ & \\
  & &  &  $ d = \frac{5}{55} $ & \\
  & &  & $ f= 0.4 $ & \\
  & &  & $ g= \frac{\ell-2}{4} $  & \\
  & &  & $ h = \frac{\ell-4}{7} $ & \\
  & &  & $ j= \frac{\ell - 6}{3} $ & \\
  & &  & $ k= \frac{1}{1 + 0.03 \ell} $ & \\ \hline
\end{tabular}
\end{table*}

\begin{table*}
\caption{ \label{InitVelPert}
The Initial Velocity Perturbations used in the runs.  All the values in
the table are dimensionless, and the distance parameter is the areal radius
in kiloparsecs $\ell = R_i$/1~kpc.
 Note that the output figures depend on the initial perturbation in both
density and velocity.
 }
\begin{tabular}{llllll}

Section & Model & Velocity perturbation & Parameters  & Graph & Output \\ \hline
\ref{inperofhom} & 1-6 &  $  \nu_{INIT}(\ell) = A \cdot ( b \arctan{c} -
 d \ell - $ & $  A = - 4 \times10^{-5} $ & Fig. \ref{beg} & Fig.
 \ref{konk},\ref{wm} \\
  & & $ f e^{-g^2} - e^{-h^2} - i e^{-j^2}) \cdot k $  &  $ b = 4  $ &  \\
  & &  &  $ c = 0.16 \ell - 2.2 $ & \\
  & &  &  $ d = \frac{5}{35} $ & \\
  & &  & $ f= 0.5 $ & \\
  & &  & $ g= \frac{\ell-7}{6} $  & \\
  & &  & $ h = \frac{\ell-9}{7} $ & \\
  & &  & $ i= 1.4 $ & \\
  & &  & $ j= \frac{\ell - 11}{3} $ & \\
  & &  & $ k= \frac{1}{1 + 0.03 \ell} $ & \\
 \hline
\ref{homovel} & 1-6 &  $  \nu = 0 $ & & & Fig. \ref{kmbp} \\
 \hline
\ref{homorho} & 1-6 &  $  \nu = \nu_{INIT}(\ell) $ & & Fig. \ref{beg} & Similar
to Figs. \ref{konk},\ref{wm} \\
 \hline
\ref{amplitude} & 1-6 & $ \nu_{AMP}(\ell) = A \cdot ( b \arctan{c}
 - d \ell - $ & $  A = - 7.5 \times10^{-4} $ & & Fig. \ref{kmbc} \\
  & & $ f e^{-g^2} - e^{-h^2} - e^{-j^2}) \cdot k $  &  $ b = 4  $ &  \\
  & &  &  $ c = 0.08 \ell - 1.1 $ & \\
  & &  &  $ d = \frac{5}{55} $ & \\
  & &  & $ f= 0.4 $ & \\
  & &  & $ g= \frac{\ell}{4} $  & \\
  & &  & $ h = \frac{\ell-2}{7} $ & \\
  & &  & $ j= \frac{\ell - 4}{3} $ & \\
  & &  & $ k= \frac{1}{1 + 0.03 \ell} $ & \\
 \hline
\ref{modelfit} & 1 &  $  \nu_{1}(\ell) = 37.5 \cdot
 \nu_{INIT}(\ell) $ & & Fig. \ref{plb} & Fig. \ref{glb},\ref{hlb},\ref{fit} \\
 \hline
\ref{modelfit} & 2,4 &  $  \nu_{2, 4}(\ell) = A \cdot ( b \arctan{c} - d
 \ell - $ & $  A = - 3.5 \times10^{-3} $ & Fig. \ref{plb} & Fig.
 \ref{glb},\ref{hlb},\ref{fit} \\
  & & $   f e^{-g^2}  - e^{-h^2}  - e^{-j^2} - $  &  $ b = 4  $ &  \\
  & & $ m e^{-n^2}) \cdot k + p $ &  $ c = 0.02 \ell - 0.02 $ & \\
  & &  &  $ d = \frac{5}{55} $ & \\
  & &  & $ f= 0.7 $ & \\
  & &  & $ g=  \ell $ & \\
  & &  & $ h = \frac{\ell-1}{7} $ & \\
  & &  & $ j = \frac{\ell-3}{3} $ & \\
  & &  & $ m= 1.225 $ & \\
  & &  & $ n= \frac{\ell - 39}{12} $ & \\
  & &  & $ k= \frac{1}{1 + 0.03 \ell} $ & \\
  & &  & $ p = 5 \times 10^{-4} $ & \\
 \hline
\ref{modelfit} & 3 &  $  \nu_{3}(\ell) = A \cdot ( b \arctan{c} - d
  \ell - $ & $  A = - 3.5 \times10^{-3} $ & Fig. \ref{plb} & Fig.
  \ref{glb},\ref{hlb},\ref{fit}  \\
  & & $   f e^{-g^2}  - e^{-h^2}  - e^{-j^2} - $  &  $ b = 4  $ &  \\
  & & $ m e^{-n^2}) \cdot k + p $ &  $ c = 0.02 \ell - 0.02 $ & \\
  & &  &  $ d = \frac{5}{55} $ & \\
  & &  & $ f= 0.7 $ & \\
  & &  & $ g=  \ell $ & \\
  & &  & $ h = \frac{\ell-1}{7} $ & \\
  & &  & $ j = \frac{\ell-3}{3} $ & \\
  & &  & $ m= 0.7 $ & \\
  & &  & $ n= \frac{\ell - 39}{12} $ & \\
  & &  & $ k= \frac{1}{1 + 0.03 \ell} $ & \\
  & &  & $ p = 5 \times 10^{-4} $ & \\
 \hline
\ref{radtt} & both &  $  \nu_{RAD}(\ell) = A \cdot ( b \arctan{c} -
 d \ell - $ & $  A = - 7.5 \times10^{-4} $ &  & Fig. \ref{grad} \\
  & & $ f e^{-g^2} - e^{-h^2} - e^{-j^2}) \cdot k $  &  $ b = 4  $ &  \\
  & &  &  $ c = 0.08 \ell - 1.1 $ & \\
  & &  &  $ d = \frac{5}{55} $ & \\
  & &  & $ f= 0.4 $ & \\
  & &  & $ g= \frac{\ell-2}{4} $  & \\
  & &  & $ h = \frac{\ell-4}{7} $ & \\
  & &  & $ j= \frac{\ell - 6}{3} $ & \\
  & &  & $ k= \frac{1}{1 + 0.03 \ell} $ & \\ \hline
 \end{tabular}
\end{table*}

\subsection{The Void Models}\label{voidmod}

\subsubsection{Initial perturbations of homogeneity }\label{inperofhom}

Because of lack of precise observational data, it is not possible to calculate
the exact profile of the initial density and velocity perturbations. From the
measurements of the microwave background radiation one can estimate only the
amplitudes of these profiles. It can be intuitively expected that the region
which in the future would become a void should have, at the initial instant, a
minimum of density and a maximum of velocity at the center.\footnote{But see the
papers by Mustapha and Hellaby (2001) and by Krasi\'nski and Hellaby (2002) --
maxima and minima can be reversed during evolution, and it is not at all
necessary that a void begins with a minimum of density at the center.}

\begin{figure}
\includegraphics[scale=0.6]{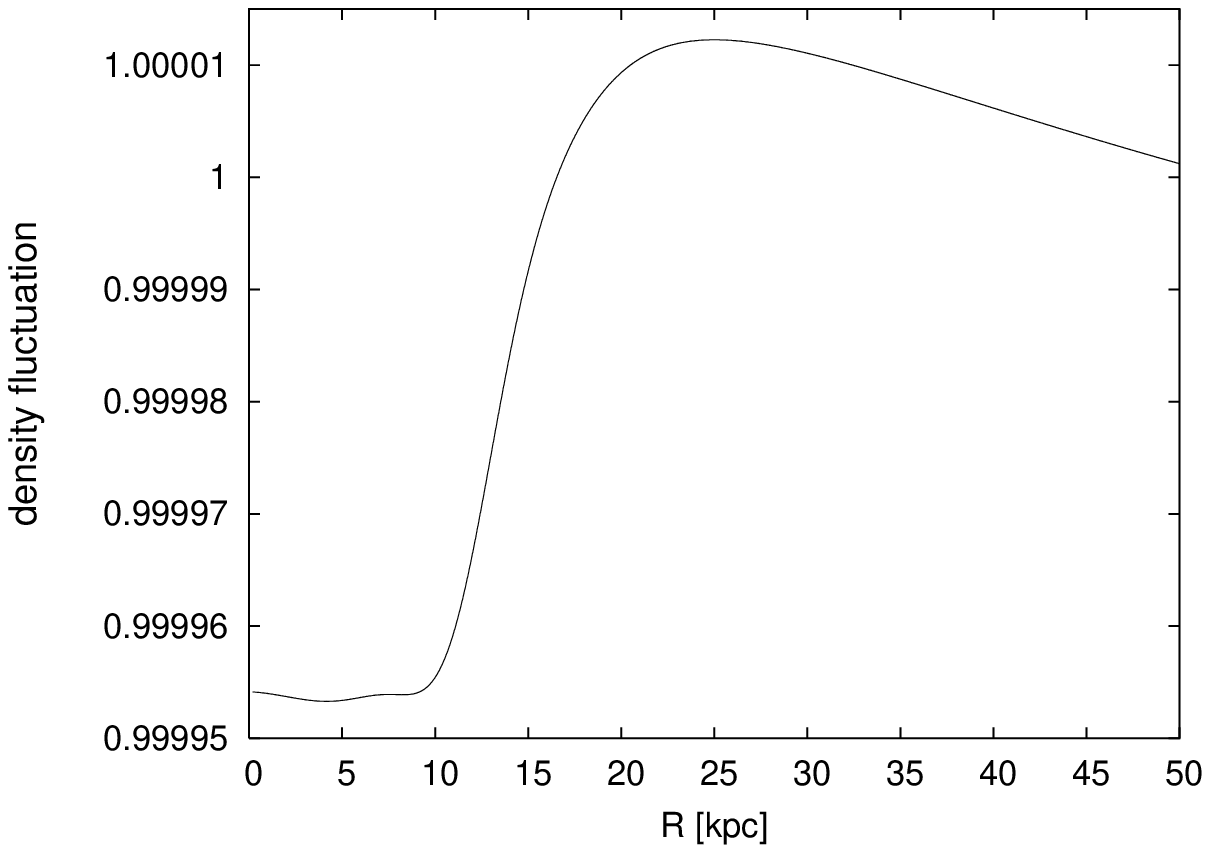}
\includegraphics[scale=0.6]{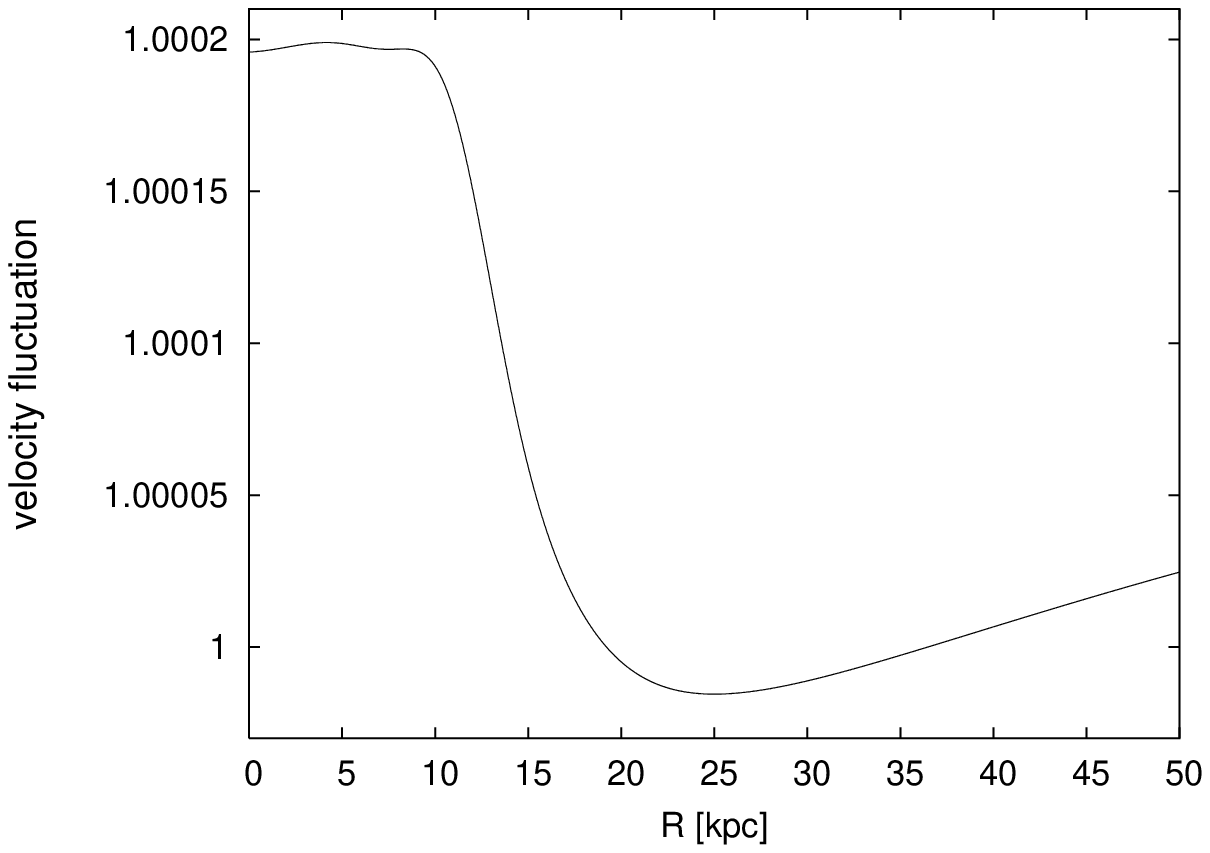}
\caption{\label{beg} The initial density (upper) and velocity (lower)
perturbations for data discussed in section \ref{inperofhom}. }
\end{figure}

The chosen initial density and velocity distributions fulfilling the above
conditions are presented in Fig. \ref{beg}. These profiles conform to the
amplitudes estimated in sec. \ref{infl}. They were defined by the functions
presented in Tables \ref{InitDenPert} and \ref{InitVelPert}. \footnote{ The form
of these initial density and velocity distributions is a result of consecutive
adjustments made in order to test the influence on void formation of the various
factors mentioned at the beginning of this section.}

\begin{figure}
\includegraphics[scale=0.7]{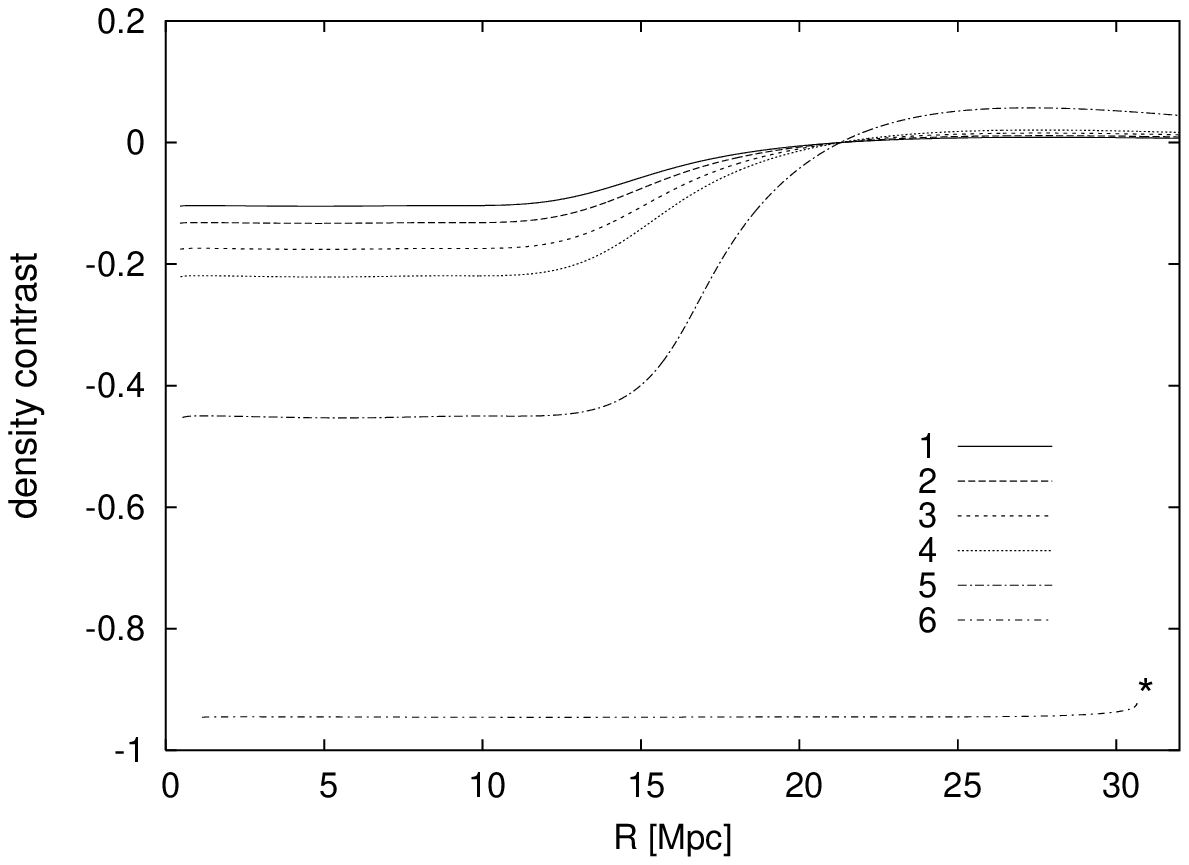}
\includegraphics[scale=0.7]{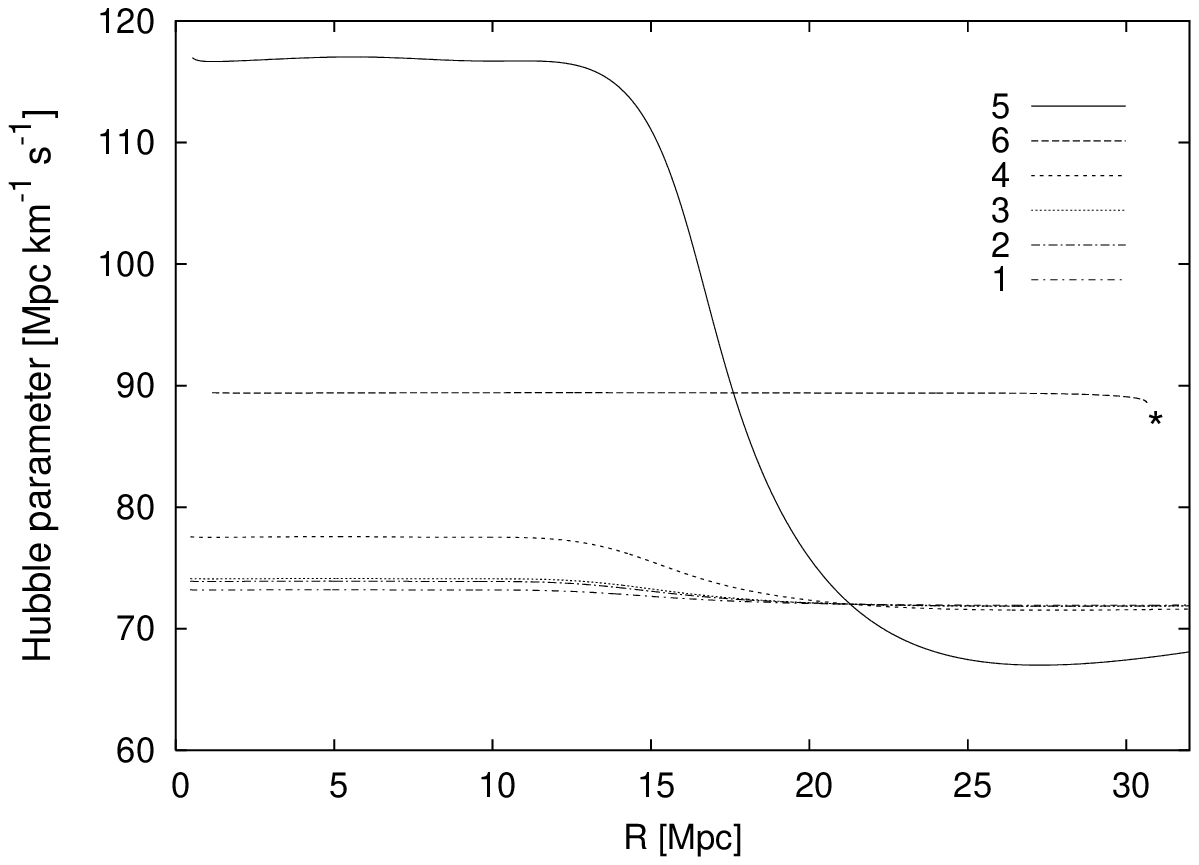}
\caption{\label{konk} The current density contrast and the Hubble
parameter for data discussed in section \ref{inperofhom}, in six
different background models: {\bf 1} -- $\Om = 0.27,~ \Ol =0$; ~{\bf 2}
-- $\Om = 0.39,~ \Ol =0$; ~{\bf 3} -- $\Om = 0.27,~ \Ol = 0.73$; ~{\bf
4} -- $\Om = 1, ~\Ol =0$; ~{\bf 5}
-- $\Om = 11, ~\Ol =0$; ~{\bf 6} -- $\Om = 0.27, ~\Ol =1.64$.
* -- shell crossing.}
\end{figure}

From the current observations, only the average density contrast is known. For
the purpose of comparing our results with the observational data (Fig.
\ref{rges}), the results shown in Fig. \ref{konk} (and also the figures showing
the final density contrast in what follows) do not present the real density
contrast, but the average one, i.e.:
 \beq
 \delta = \frac{ < \rho> }{\overline{\rho}} - 1,
 \label{dcon}
 \eeq
 \noindent
where $\overline{\rho}$ is the present background density:
 \beq
 < \rho> = \frac{ 3 M c^2}{ 4 \pi G R^3}.
 \eeq

Unfortunately there are no astronomical data for the current velocity
distribution in the void. It is more practical to measure the expansion
rate by the equivalent of the Hubble parameter
$(R_{,t}/R)$. The results are presented in Fig. \ref{konk}.

\begin{figure}
\includegraphics[scale=0.7]{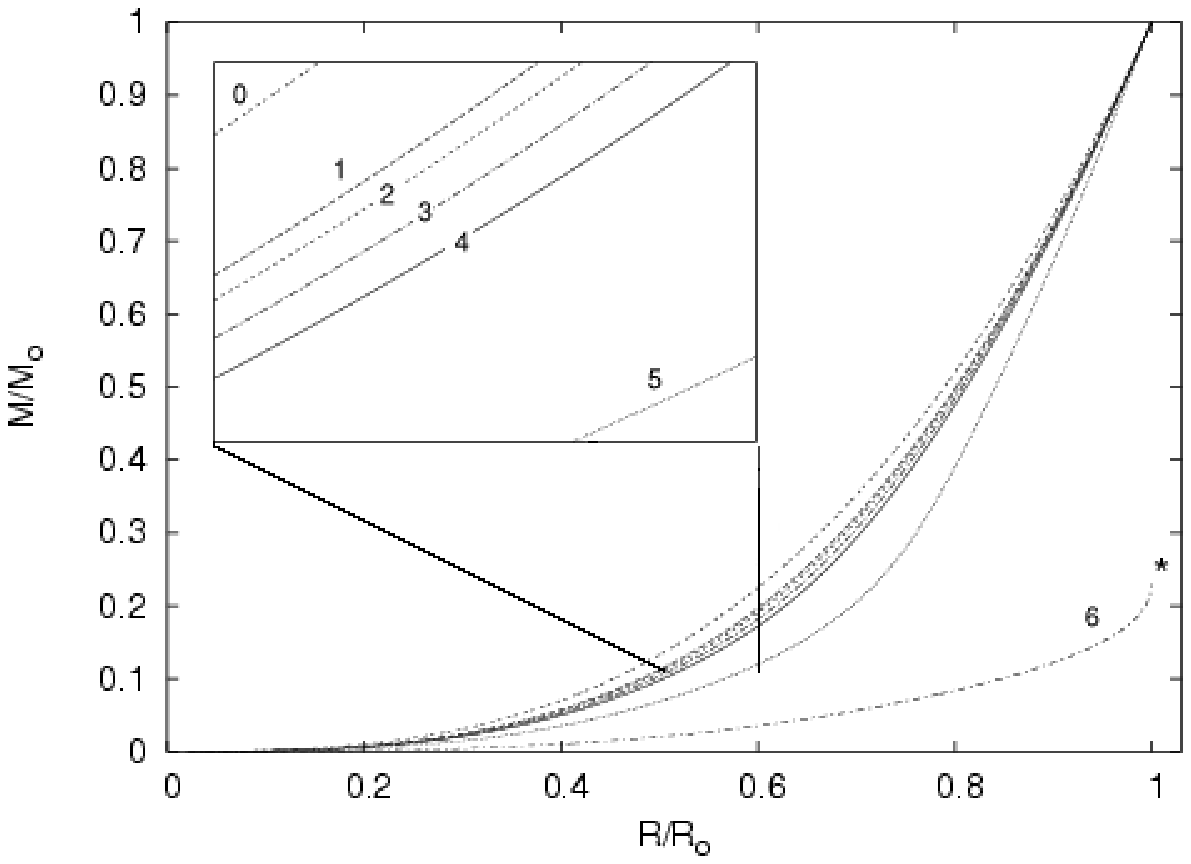}
\caption{\label{wm} The mass re-distribution for data discussed in section
\ref{inperofhom}. $R_0$ is the smallest $R$ values at which the density takes
the background value. $M_0$ is the mass inside the shell of areal radius $R_0$.
{\bf 0 } -- the initial condition; other labels as in Fig. \ref{konk}.}
\end{figure}

Fig. \ref{wm} shows the re-distribution of mass resulting from void formation.
Curve {\bf O} is the function $M(R)$ at the initial time $t = t_{ls}$ (i.e.
$M(r)$ versus $R(r, t_{ls})$), and the other curves show the calculated $M(R)$
at the present time, $t = t_0$, for different background models.

Curve {\bf 6} in Fig. \ref{konk} is truncated. At the cutoff point a shell
crossing occurs -- the inner shells catch up with the outer shells. This results
in a singularity that probably does not occur in the real Universe. Before it
happens, the gradient of pressure would become significant and the \lt model
would become invalid.

From the figures presented above some initial conclusions can be made. One can
say that the current depth of the void depends of the amount of mass
re-distribution, i.e. the mass outflow. This outflow depends on the expansion
rate and on the age of the void. The major influence on the structure formation
is from the shell expansion.

In models {\bf 4} and {\bf 5} the expansion rate is bigger than in models {\bf 1
-- 3} and the final density contrast in models {\bf 4} and {\bf 5} is much more
negative than in models {\bf 1 -- 3}, even though the age of the Universe in
these models is much lower.

Unfortunately, models {\bf 1 -- 5} cannot recover the observed density contrast
of today's voids. The proper depth can be obtained only in model {\bf 6}, where
the age of the Universe is significantly larger.

The age of the void (in limits estimated by the various astronomical
observations) is of lesser importance, compared to the expansion rate. In models
{\bf 2} and {\bf 3} the expansion of the shells is similar and the final density
contrast in model {\bf 3} is lower due to the time available for evolution being
by $2 \times 10^9$ years longer.

We now check how big the influence of the initial shape of the
density and velocity perturbations is.

\subsubsection{Homogeneous velocity profile }\label{homovel}

For this case, the initial density profile $\delta$ is as in Fig. \ref{beg},
while the initial velocity profile is $\nu = 0$. The explicit formulae for these
profiles are presented in Tables \ref{InitDenPert} and \ref{InitVelPert}.

\begin{figure}
\includegraphics[scale=0.7]{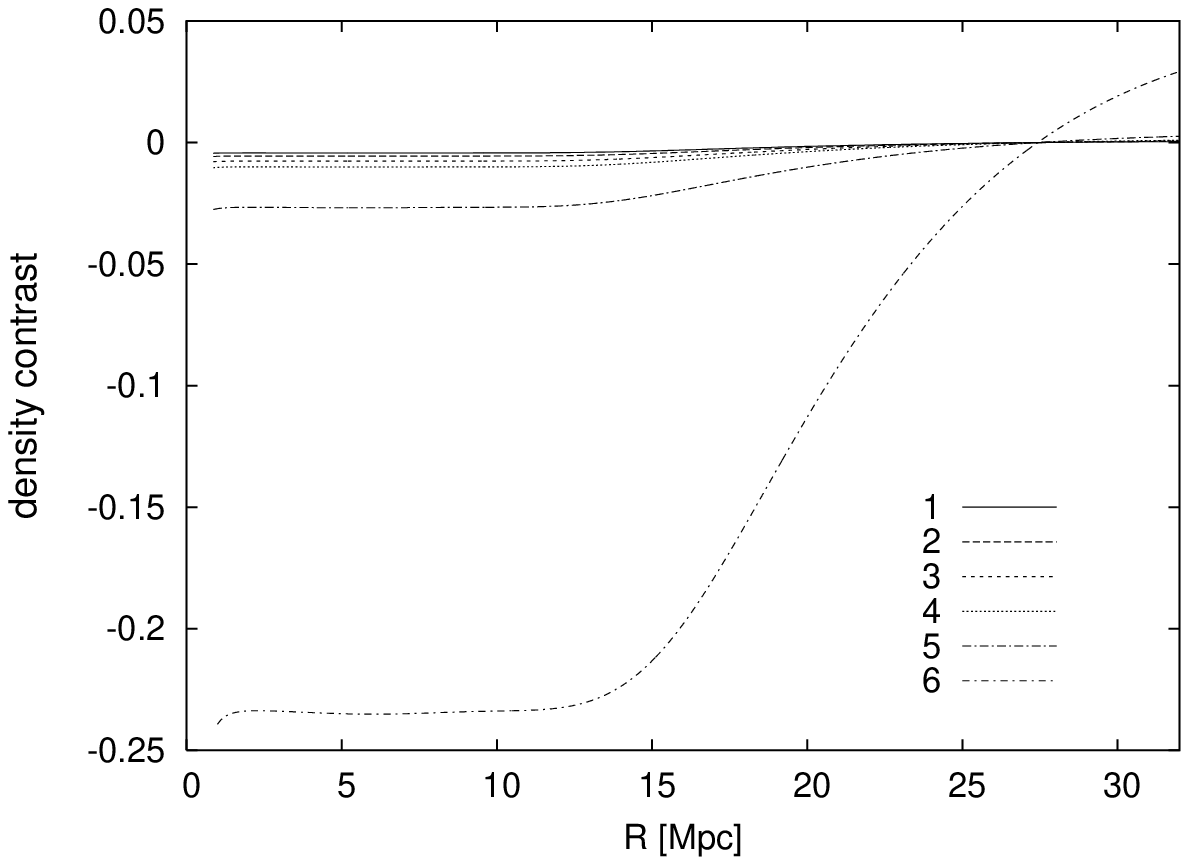}
\includegraphics[scale=0.7]{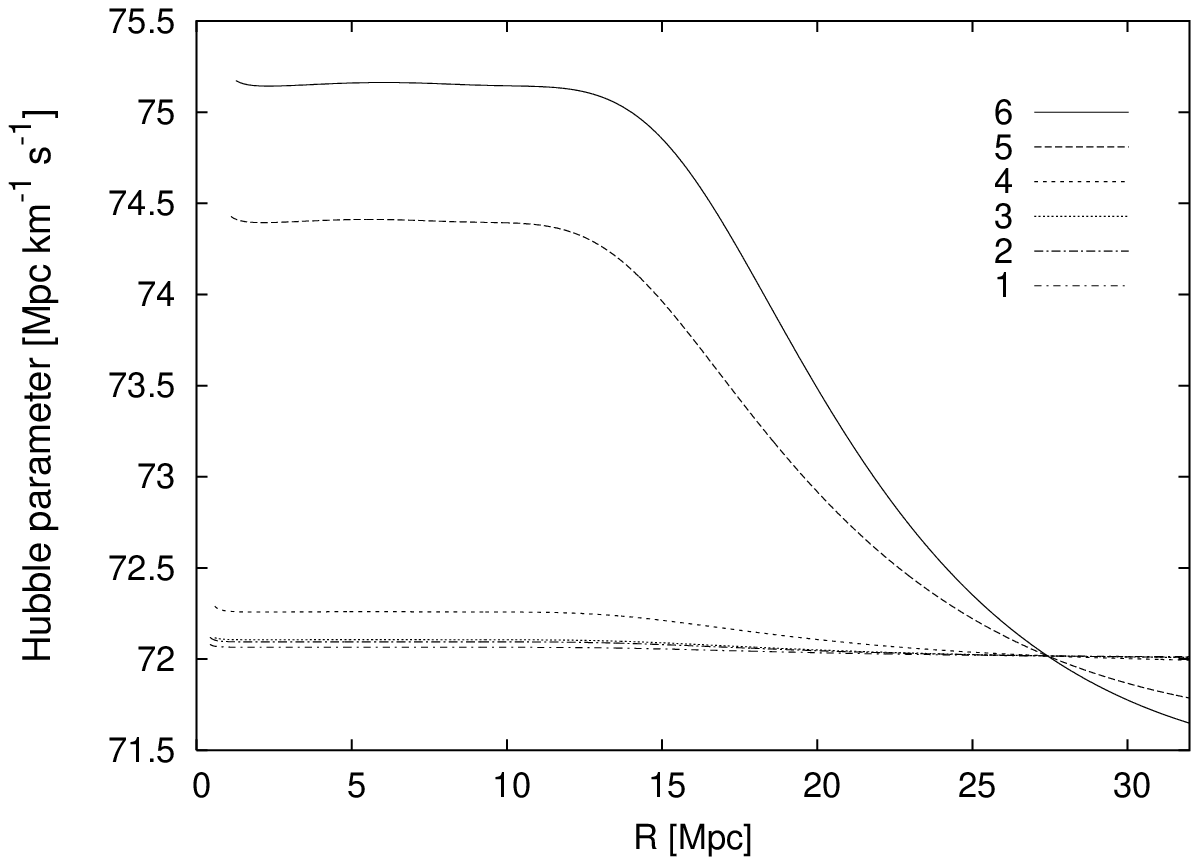}
\caption{\label{kmbp} The current density contrast and the Hubble parameter for
the flat initial velocity profile in different background models. Labels as in
Fig. \ref{konk}}
\end{figure}

The results shown in Fig. \ref{kmbp} seem to be surprising. The mass
re-distribution is almost the same in all the models (except model {\bf 6}), the
diameters are similar, but the current density profiles are different. In this
case, a second factor responsible for the void formation is seen, which is the
faster expansion rate. The faster expansion of the void compared to the
homogeneous background causes that the difference between the density in the
central regions of the void and the density in the background  increases with
time, even when the mass of the shell inside the region of $R(t,r)$ is not
changing very much. Consequently, in the models with higher expansion  rate the
density contrast is most negative.

\subsubsection{Homogeneous density profile}\label{homorho}

In contrast to the above, we now set the initial density profile to $\delta
= 0$, while the initial velocity profile, $\nu$, is as in Fig. \ref{beg}.
The explicit formulae for the profiles are presented in Tables
\ref{InitDenPert} and \ref{InitVelPert}.

The final results are not very different from the one shown in Fig. \ref{konk}.
This proves that the velocity distribution in the formation of voids is very
significant, while the density distribution is of lesser importance.

\subsubsection{Amplitude}\label{amplitude}

\begin{figure}
\includegraphics[scale=0.7]{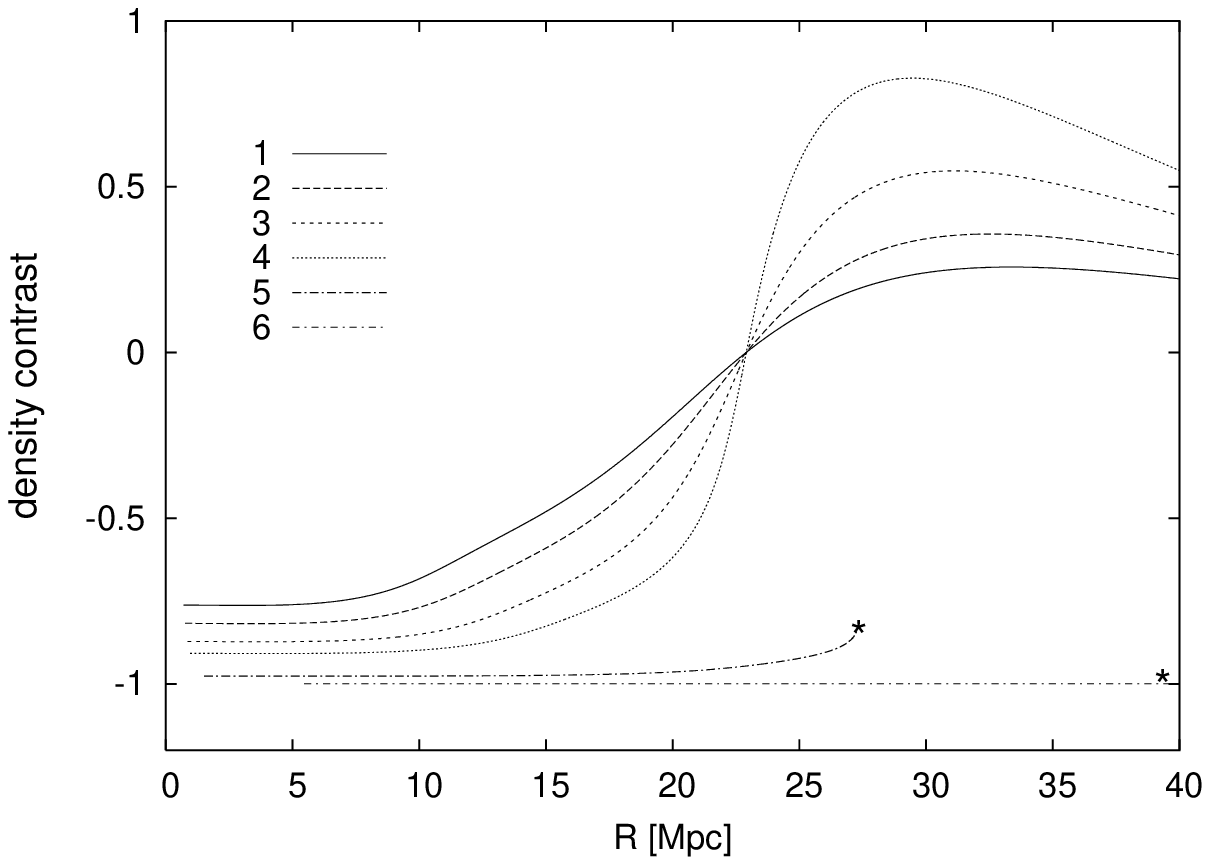}
\includegraphics[scale=0.7]{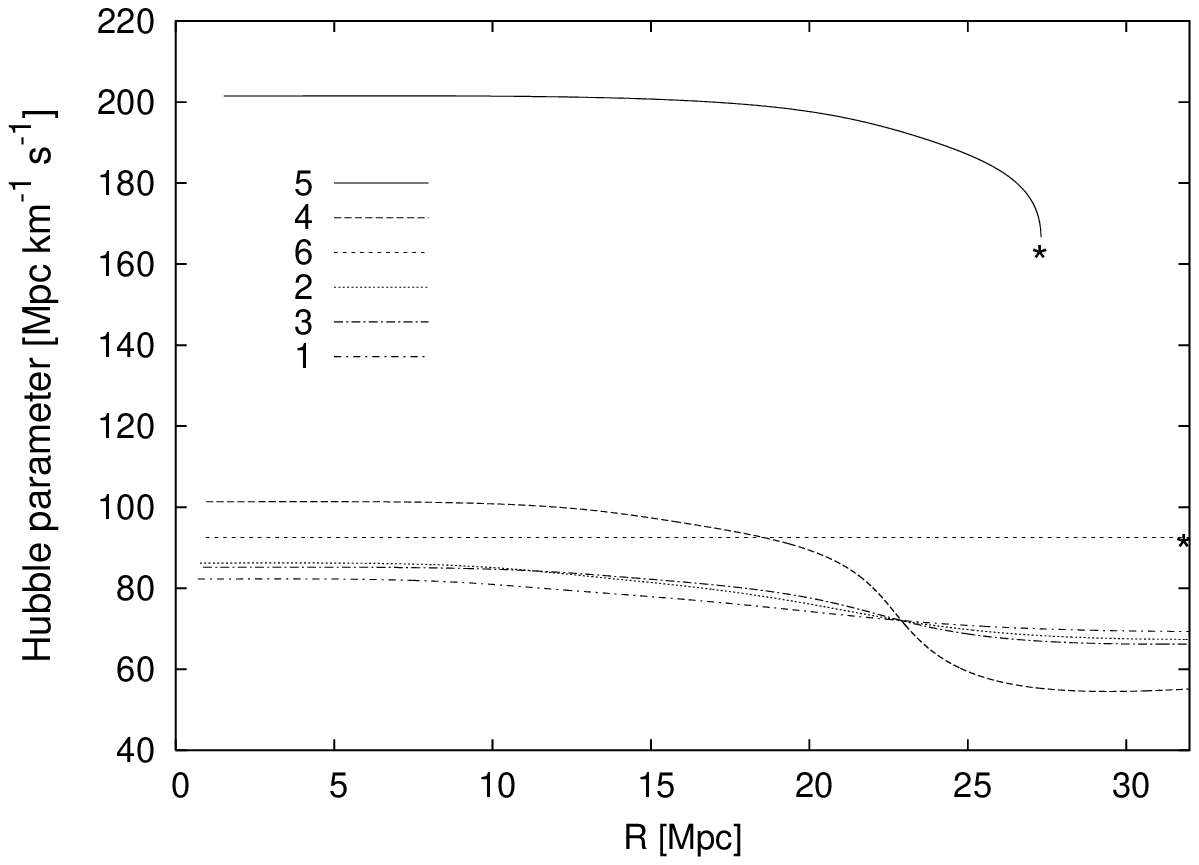}
\caption{The current density contrast and the Hubble parameter for the initial
data with higher amplitude of density and velocity perturbation, in six
different backgrounds models. Labels as in Fig. \ref{konk}. \label{kmbc}}
\end{figure}

In this subsection, the amplitude of the initial fluctuations is increased
compared to the one used in subsection \ref{inperofhom} and is  $4 \cdot
10^{-3}$. The profiles of the initial perturbations are presented in Tables 3
and 4, and the final results in Fig. \ref{kmbc}.

The increased amplitude of the initial perturbations results in a void with a
higher negative density contrast. To obtain a density contrast near $\delta \sim
-0.9$ we needed to increase the amplitude of the initial density profile more
than 70 times, and the amplitude of the velocity profile 20 times, compared to
the values estimated from CMB fluctuations. Even so, the value $\delta = -0.94 $
of the density contrast in the void was not reached, except in the two
non-realistic background models. In model {\bf 4} the minimum value is $-0.908$
and in model {\bf 3} it is $-0.873$. Unfortunately,  increasing the amplitude
leads to a shell crossing singularity in some models.

\subsubsection{Observation and a model - a cross check}\label{modelfit}

In the previous sections we had problems generating voids from small initial
density and velocity fluctuations. The only alternative was to use a background
model with an extremely large age of the Universe (inconsistent with limits
estimated by the various astronomical observations).

In this subsection, we will try to choose the initial profiles that lead to the
best fit with observational data. We focus only on one background model,
preferred by the astronomical observations, which is $\Om= 0.27$ ~and~ $\Ol =
0.73$. It should be noted that the difference between void evolution in this
model and in the Einstein-de Sitter model is not big, so the problem of void
formation is in the initial conditions rather than in the background model.

So far none of the results obtained have recovered the observational data,
presented in Fig. \ref{rges}. This figure presents the density profile of the
void with very smooth edges. Unfortunately this profile does not reach to
regions where the  density is higher than the background density, and where the
superclusters of galaxies would be found. The mean value of the density contrast
inside the void is $ \delta \sim -0.94$. The density contrasts of the voids
obtained so far have been too shallow or had steep edges.

\begin{figure}
\includegraphics[scale=0.7]{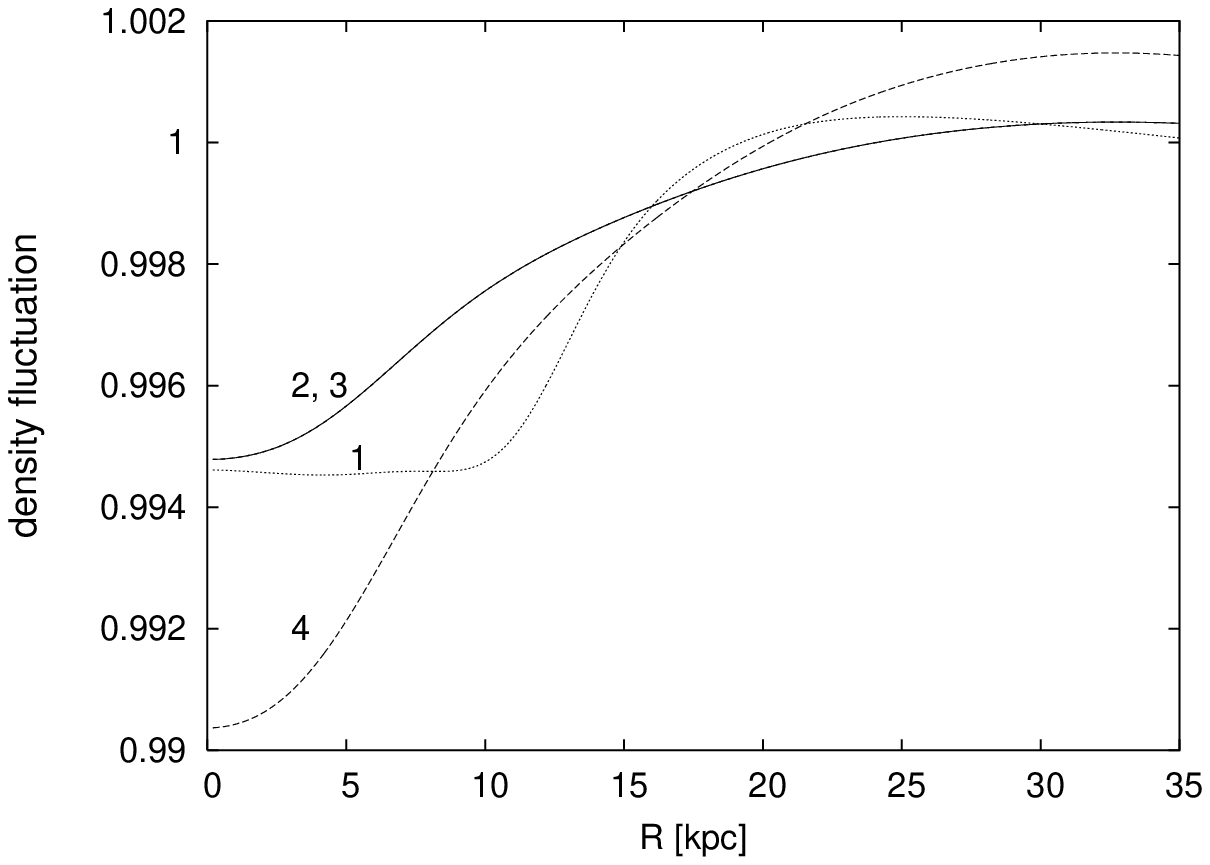}
\includegraphics[scale=0.7]{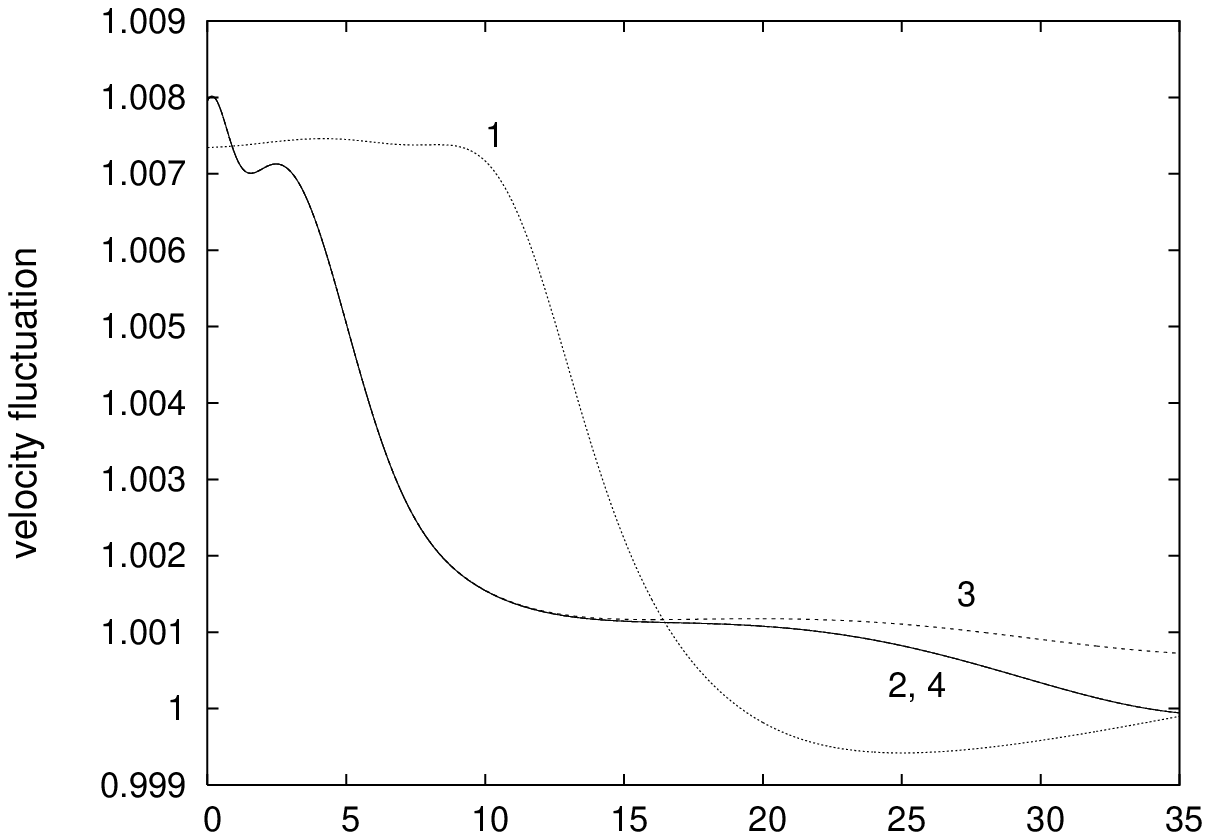}
\caption{\label{plb} The initial density (upper) and velocity (lower)
perturbations for the results presented in Figs. \ref{glb} and
\ref{hlb}.}
\end{figure}

\begin{figure}
\includegraphics[scale=0.7]{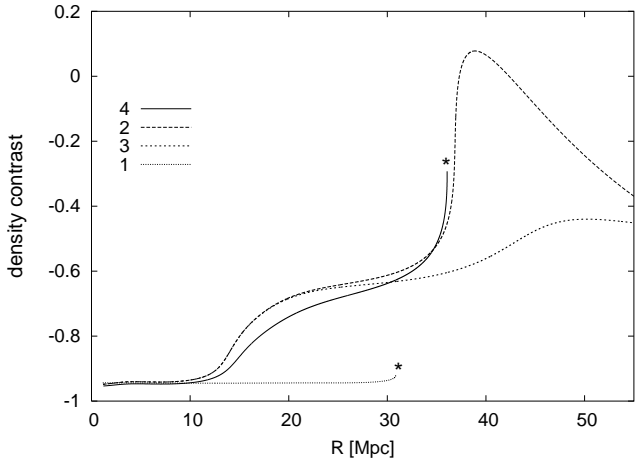}
\caption{\label{glb} The current density contrast for the models
discussed in section \ref{modelfit}.}
\end{figure}

\begin{figure}
\includegraphics[scale=0.7]{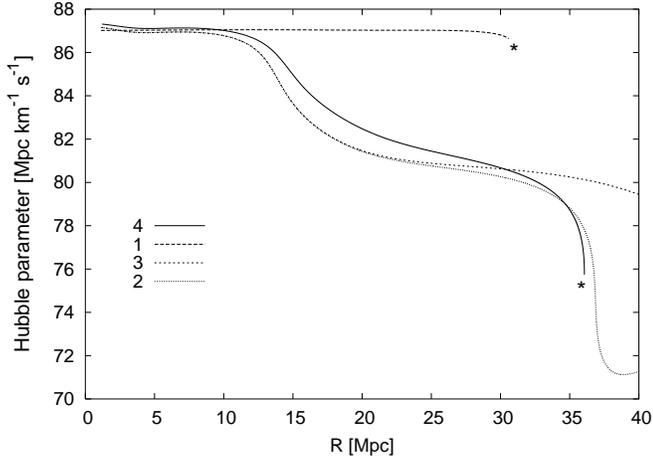}
\caption{\label{hlb} The Hubble parameter for the models discussed in
section \ref{modelfit}.}
\end{figure}

The initial fluctuations are presented in Fig. \ref{plb}.  The profiles are
presented in Tables \ref{InitDenPert} and \ref{InitVelPert}.  The results are
shown in Fig. \ref{glb} and \ref{hlb}. Model {\bf 2} has both proper density
contrast and smooth edges. The conclusion from numerical experiments with
different shapes of the initial profiles is that a model of a void consistent
with observational data (with the value of density contrast less then $ \delta =
-0.94$, smooth edges and high density in the surrounding regions) is very hard
to obtain within the \lt model, without the occurrence of a shell crossing
singularity. The final  state of model {\bf 2} was very close to this
singularity and in model {\bf 4} a shell crossing occurred.

\begin{figure}
\includegraphics[scale=0.65]{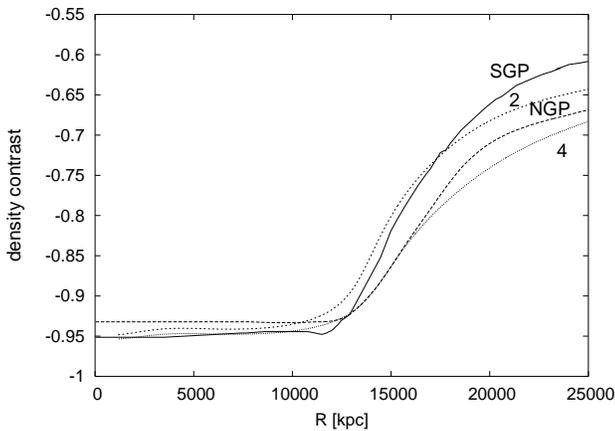}
\caption{\label{fit} Comparison of the curves from Fig. \ref{glb} ({\bf 2,4})
with the observed density contrast from Hoyle and Vogeley (2004) ({\bf SGP} and
{\bf NGP}). }
\end{figure}

The main factor responsible for void formation is the velocity perturbation,
with an amplitude of $\sim 8 \cdot 10^{-3}$, near the centre, and dropping below
zero in the outer regions (model {\bf 3} did not fulfill this condition). The
density fluctuation is of lesser importance. The models {\bf 1 -- 3} had the
same initial density fluctuations and in model {\bf 4} the amplitude was two
times greater. In spite of these differences, the final results differ only in
shape, and not in the depth of the final density contrast.

In simulations with higher values of background density the initial velocity and
density perturbation needed to obtain similar final results are smaller (see
subsection \ref{amplitude}).

\section{Evolution}\label{evol}

Let us take a closer look at the evolution of model {\bf 2}.

\begin{figure}
\includegraphics[scale=0.65]{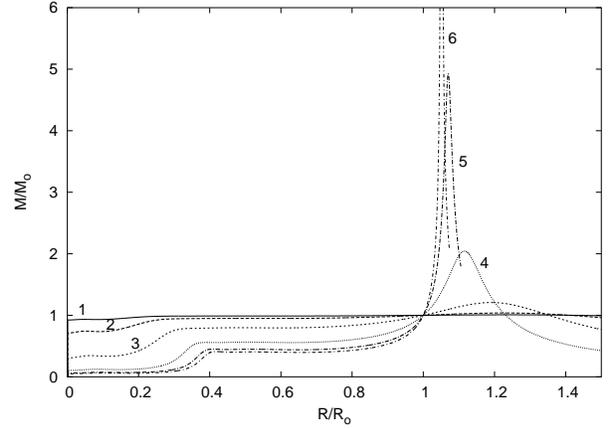}
 \caption{\label{ewol}
The evolution of the density distribution in a void for the age of the Universe:
10 (No. {\bf 1}), 100 (No. {\bf 2}) million years and 1 (No. {\bf 3}), 5 (No.
{\bf 4}), 10 (No. {\bf 5}) billion years after the Big Bang, and the current
(No. {\bf 6}) profile of density distribution for the model discussed in section
\ref{evol}. $R_0$ is the point at which the density takes the background value
for the first time.}
\end{figure}

Fig. \ref{ewol} shows the density distribution in eight
different moments of time. In this section the figures do not present
the density contrast, but the real density distribution calculated from
eq. (\ref{LTden}).

\begin{figure}
\includegraphics[scale=0.65]{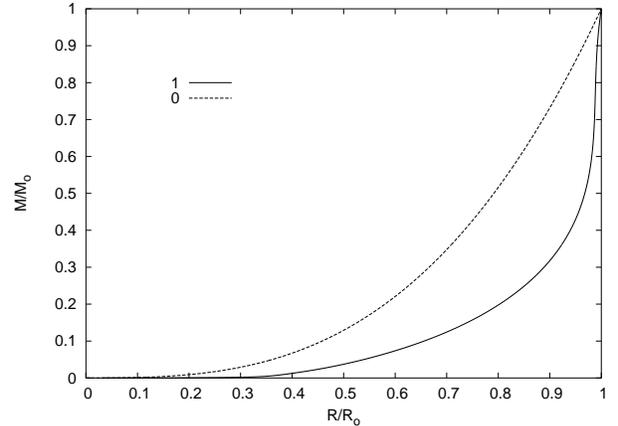}
\caption{\label{mevo} The mass re-distribution for data discussed in section
\ref{evol}. $R_0$ is the point at which the density takes the background value
for the first time. $M_0$ is the mass inside the shell of areal radius $R_0$.
Curve {\bf O} is the function $M(R)$ with $R$ and $R_0$ taken at $t_1$, curve
{\bf 1} is the function $M(R)$ with $R$ and $R_0$ taken at time $t_2$.
  }
\end{figure}

\begin{figure}
\includegraphics[scale=0.65]{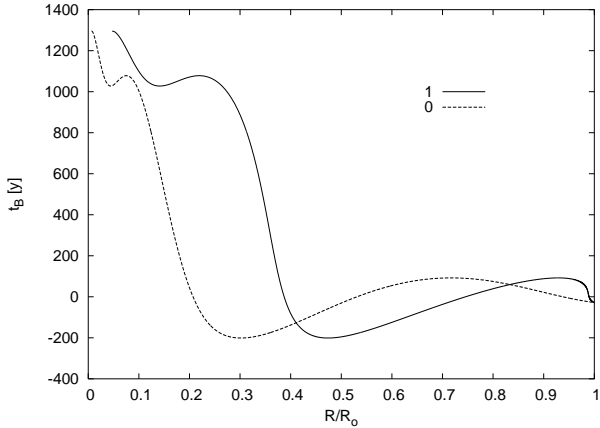}
\caption{\label{tevo}
 The Bang time  $t_B$ as a function of $R(t_1, r)$ (curve {\bf O}) and of
$R(t_2, r)$ (curve {\bf 1}) for the model discussed in section \ref{evol}.
  }
\end{figure}

\begin{figure}
\includegraphics[scale=0.65]{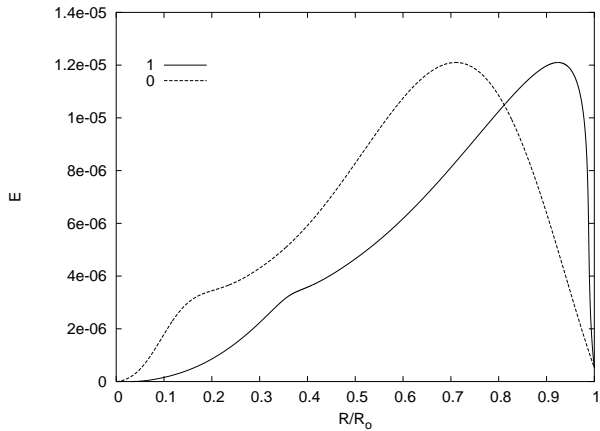}
\caption{\label{eevo}
 The energy function $E$ as a function of $R(t_1, r)$ (curve {\bf O}) and
of $R(t_2, r)$ (curve {\bf 1}) for the model discussed in section
\ref{evol}.
   }
\end{figure}

Figs. \ref{mevo} -- \ref{eevo} show respectively the functions $M(R)$,
$t_B(R)$ and $E(R)$, where $R$ is the areal radius at the initial
instant (curves {\bf O}) and at the final instant (curves {\bf 1}). The
pictures demonstrate the evolution of the structure: in the expanding
void mass moves outwards.

As one can see from Fig. \ref{tevo}, the difference in the value of the Bang
time function between outer and central regions of the void is of the order of
hundreds of years. This is negligible compared to the age of the Universe which
at the moment of last scattering was of the order of $10^5$ years.

\section{Limitations on the initial conditions }

An exact reconstruction of the initial conditions that held at the time
of decoupling in the region that would become a void is not possible due
to the lack of precise data. Nowadays the only way to reconstruct the
initial conditions is the theoretical approach. The linear theory is
often used for this purpose because of its simplicity.

\subsection{Initial conditions in the linear theory }

In linear theory the shape of the density perturbation is constant in
time. Only the amplitude is changing, according to the formula:
 \beq
\delta (t) = D \delta_0,
 \eeq
where $\delta_0$ is the current value and D is the linear growth factor.
In the Einstein-de Sitter Universe it is equal to:
 \beq
D = \frac{1}{1+z},
 \eeq
and in the general case:
 \beq
D(a)\ = \frac{5 \Omega_{M}}{2a f(a)} \int\limits_{0}^{a} f^3(a') da',
 \eeq
where
 \beq
f(a) = \frac{1}{\sqrt{1 + \Omega_M(\frac{1}{a} -1) +
\Omega_{\Lambda}(a^2 -1)}},
 \eeq
while $a= 1/(1+z)$. For large redshifts in any model:
 \beq
D(a) \to \frac{1}{1+z}.
 \eeq

In the linear regime in the spherically symmetric case, the peculiar
velocity of the density perturbation is given by the formula (Peebles
1993, pages: 115-116):
 \beq
v(r) = - \frac{1}{3} f H S r \delta,
 \eeq
where $H$ is the Hubble constant, $S$ is the scale factor and $f$ is the
velocity factor. For large redshifts in any cosmological model this
asymptotically becomes unity. Since:
 \beq
 H = \frac{S_{,t}}{S},~~~~~~~~R_{,t} = r S_{,t},
 \eeq
the velocity fluctuations are determined by the formula:
 \beq
\nu = -\frac{1}{3} \delta
 \eeq
That implies that to calculate the evolution of the void one needs the
initial density and velocity perturbations with the amplitudes:
 \beq
 \delta \approx 9 \cdot
10^{-4}, ~~~~~~~~~~~~ \nu \approx 3 \cdot 10^{-4}
 \eeq
These results are more than 10 times smaller than in the \lt model, where
the amplitude needed is $8 \cdot 10^{-3}$

Since the current evolution is not linear, one cannot use the linear theory to
predict the initial conditions in the young Universe. However the initial
conditions could be obtained with the help of the \lt model, since it is an
exact solution of the Einstein filed equations. There still remains the question
about its accuracy, since the \lt Universe is a very simple model of a void. One
can try to answer this question by a careful look at the restrictions of the \lt
model.

\subsection{Main limitations of the \lt model  }

The \lt model is spherically symmetric, so it cannot take rotation into
account. The density and velocity distributions can only depend on the
radial coordinate. However, astronomical observations show that for voids
these conditions are fulfilled with a satisfactory accuracy.

The next limitation is the dust energy-momentum tensor, which means that
the pressure of matter and of radiation are neglected. At the present
epoch this is correct, but one can ask about the error at the moment of
the last scattering implied by this assumption. For the purpose of
estimating that error, a general Friedmann -- Lemaitre -- Robertson --
Walker (FLRW) model will be used.

\begin{enumerate}
\item
The pressure: \\
The equation of state for the perfect gas is:
\beq
 p = \frac{ \rho k_B T}{ \mu m_{H}},
\label{pg} \eeq \noindent where $k_B$ is the Boltzmann constant and
$m_H$ is the proton mass. Because of lack of data about the nature of
dark matter, let us assume that the mean molecule weight $\mu = 1$. In
FLRW models, $T = T_0 (1+z)$, where $T_0$ is the present value of the
background temperature. Substituting the numerical values we obtain that
the ratio of the pressure to the density of matter is:
\beq
 \frac{p}{\rho c^2} \approx 2.75 \cdot 10^{-10}.
 \eeq

\item
The radiation energy density:  \\

The expression for the density of the radiation energy is:
 \beq
\epsilon_{rad} = a T^4 = a T_o^4 (1+z)^4, \label{eprad}
 \eeq
where $a= \frac{4 \sigma}{c}$, and $ \sigma$ is Stefan-Boltzmann
constant. The density of matter in the FLRW models is:
 \beq \rho =
\rho_o (1+z)^3.
 \eeq
If we assume that that the present value of density is equal to the
critical density in the flat FLRW model, then the ratio of the radiation
energy density to the matter energy density will be:
 \beq
\frac{\epsilon_{rad}}{\epsilon_{mat}} = \frac{a T_o^4}{\rho_o c^2} (1 + z) \approx
0.054
\label{errad}
 \eeq

\end{enumerate}

From the above one can see that the gas pressure is negligible. Ignoring the
radiation one makes a small error which at first seems to be of little
importance. However, from the above simulations with different background
models, it can be seen that if the initial energy density is higher then the
final contrast of matter density is deeper. Unfortunately, this leads to a
shorter age of the Universe. To obtain the proper age of the Universe, a higher
value of the cosmological constant is needed.

Radiation avoids this problem because during the evolution it becomes
less and less significant and the age of the Universe does not change so
much. Is it possible then that a model of void formation that includes
radiation could predict the voids observed today starting from smaller
initial perturbations?

\subsubsection{Radiation}\label{radtt}

The Einstein field equations for the spherically symmetric perfect fluid
distribution can be reduced to the two following equations (Lemaitre
1933):
\begin{equation}
 \kappa R^2 R_{,r} \rho c^2 = 2M_{,r},
\label{mr}
\end{equation}
\begin{equation}\label{emte}
 \kappa R^2 R_{,t} p = -2M_{,t},
\label{mt}
\end{equation}
where $\rho$ is the energy density, while $p$ is the pressure. $M$ is
defined by the formula:
\begin{equation}\label{emdef}
 2M(r,t)= R(r,t){R_{,t}}^2(r,t) - 2E(r)R(r,t) - \frac{1}{3} \Lambda R^3(r,t).
\label{1c}
\end{equation}
As in the case without radiation, $M c^2/G$ is equal to the mass inside the
shell of radial coordinate $ r$. In this case, the mass is not constant in time
and in the expanding universe it decreases.

From the equations of motion ${T^{\alpha \beta}}_{; \beta} = 0$ we obtain
\begin{eqnarray}
 p_{,r} &=& 0, \nonumber \\
 p_{, \theta} &=& 0, \nonumber \\
 p_{, \phi} &=& 0,
\label{prad}
\end{eqnarray}
\begin{equation}
 \rho_{,t} + \left( \rho + \frac{p}{c^2} \right) \left(
 \frac{2R_{,t}}{R} + \frac{R_{,rt}}{R_{,r}} \right) = 0.
\label{1e}
\end{equation}
Equations (\ref{prad}) require that pressure can only be a function of time. If
inhomogeneous radiation should be included in the model, the metric form should
be changed --- the $ g_{00}$ component should be a function of the radial
coordinate. This would require finding a new exact solution of the Einstein
equations.

However, it is instructive to know what changes are caused by homogeneous
radiation. The time dependence is the same as in the Friedmann models:
\[ \epsilon_{rad} = a T^4 = \epsilon_{dec} \left( \frac{t_{dec}}{t} \right)^{8/3} \]
\[ p_{rad} = \frac{1}{3} \epsilon_{rad}. \]
For the purpose of comparing with the previous results, the background model is
the flat Friedmann model with $\Lambda = 0$. The initial density and velocity
distributions are presented in Table \ref{InitDenPert} and \ref{InitVelPert}.

The computer algorithm implemented to do the calculation was similar to
the one used in models without radiation. The only difference is that
instead of solving one equation (eq. (\ref{LTevol})), the
second-order Runge--Kutta method was used to solve simultaneously eqs.
(\ref{emte}) and (\ref{emdef}).

\begin{figure}
    \begin{center}
    \includegraphics[scale=0.65]{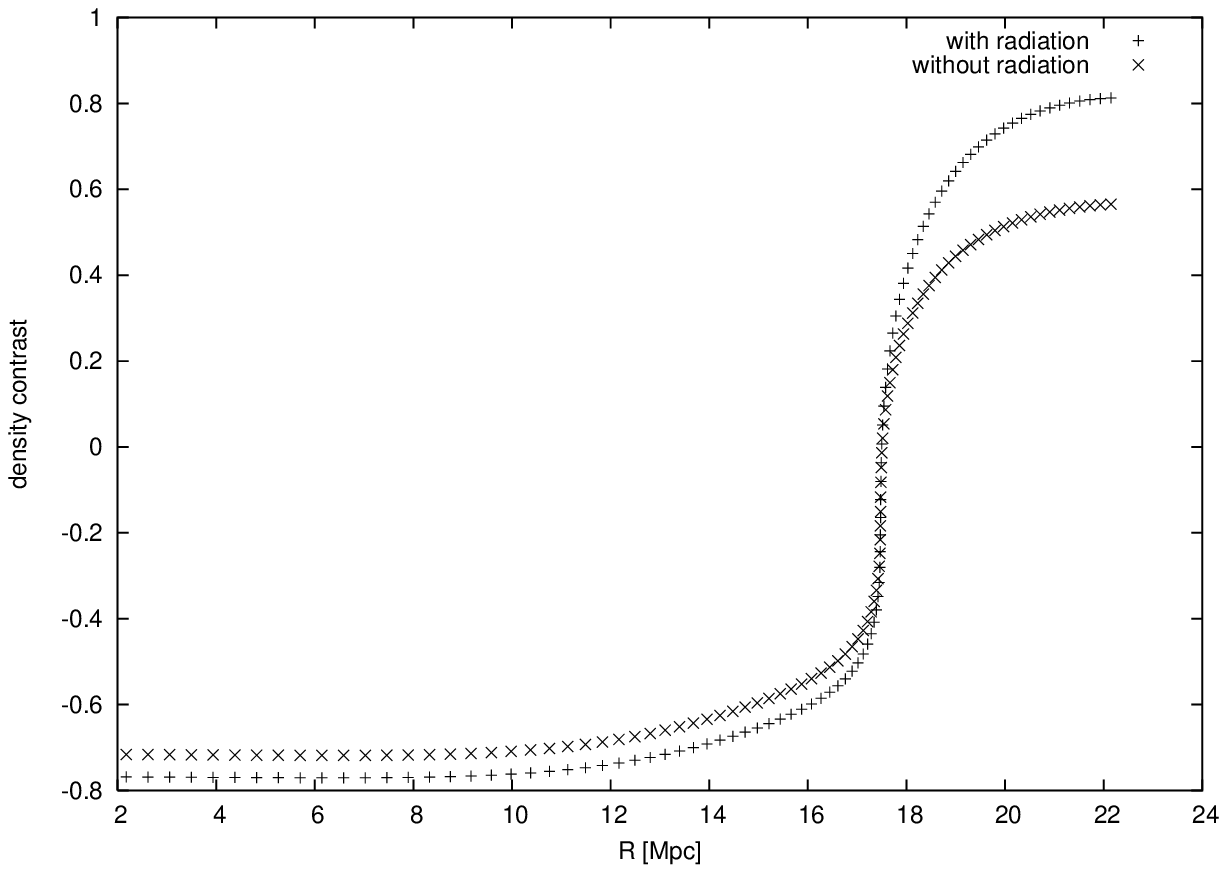}
    \includegraphics[scale=0.65]{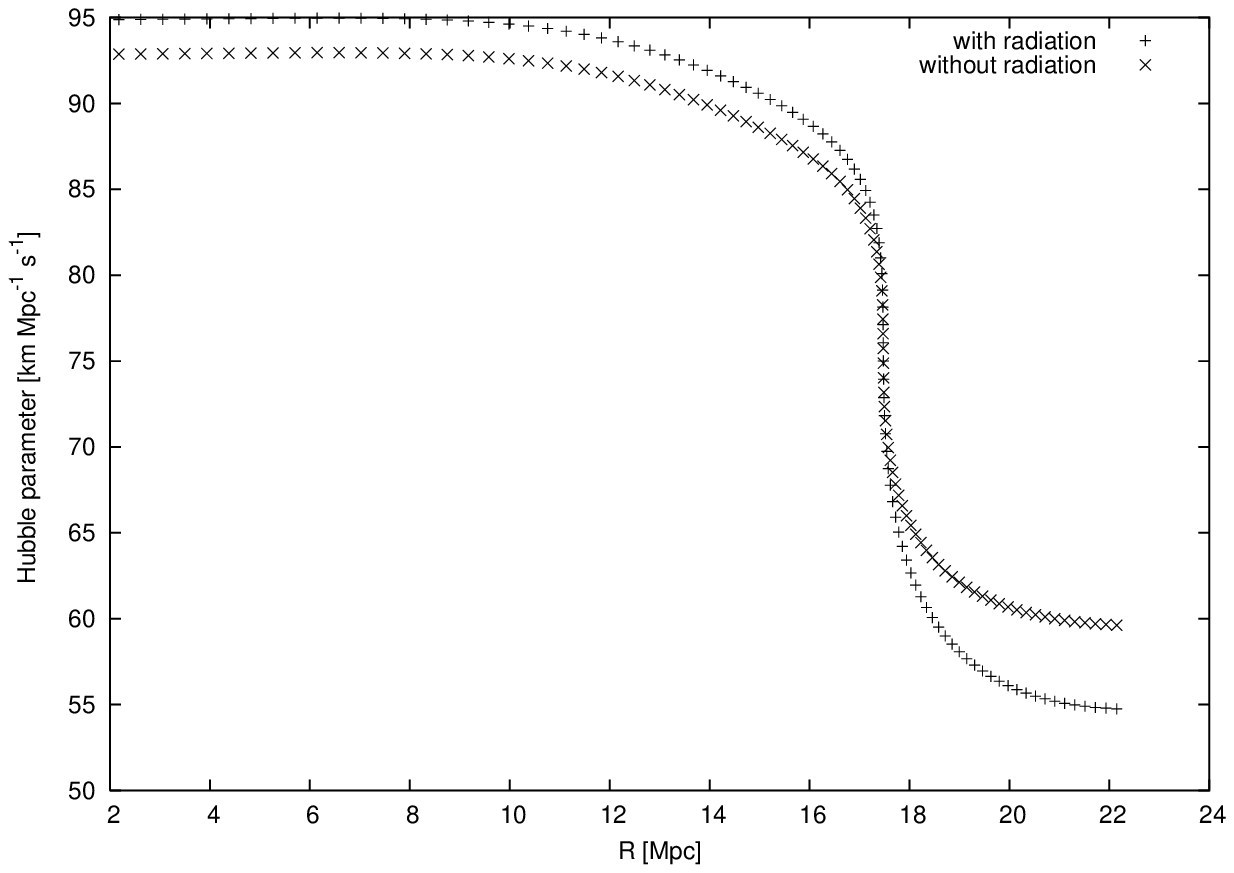}
    \caption{ \label{grad}
The final density contrast and the Hubble parameter in the model with (+)
and without (x) the radiation.}
    \end{center}
  \end{figure}

Fig. \ref{grad} shows the comparison between the models with and without
radiation. The final density contrast in the model with radiation is a
little lower inside the void, but higher at the edge compared to the
model without radiation.

As one can see, homogeneous radiation is not of great importance in the
formation of voids. The difference from the model without radiation is not big
because, in spite of higher energy density, as one can see from eq. (\ref{mt}),
the mass of the shell is decreasing. The decrease of the mass implies, via eq.
(\ref{emdef}) with $\Lambda = 0$, that the velocity of the shell is also
decreasing with time. Consequently the evolution of the structure slows down and
the final results do not show significant differences compared to the model
without radiation. These results suggest that the error estimated with eq.
(\ref{errad}) is indeed of little importance and even the model with
inhomogeneous radiation should not lead to significant differences in the final
results.

\section{Conclusion}

The main aim of this paper was to produce a non-linear model of void formation,
starting from small initial density and velocity fluctuations that existed at
last scattering, and to investigate what factors are necessary to reproduce
current observations.

\begin{itemize}

 \item   In the numerical experiments that we carried out, the perturbations of
that were needed to form a realistic present day void had to have a density
amplitude of $\delta \rho/\rho \approx 5 \cdot 10^{-3}$ 
and a velocity amplitude $\delta V/V \approx 8 \cdot 10^{-3}$ (in the
model with $\Om = 0.27$ and $\Ol = 0.73$).

 \item   It was found that density perturbations are of lesser significance than
velocity perturbations in the process of void formation.

 \item   There was no significant difference between the evolution of the void
in the model with and without the cosmological constant. In this latter
case the amplitude of the initial velocity fluctuation needed for this
purpose was a little smaller.

 \item   The existence of voids is closely related to the existence of regions of
higher density surrounding the voids. In our simulations, there were
problems obtaining reasonable profiles for these high density regions,
since shell crossing singularities tended to occur.  The \lt model breaks
down at shell crossings, so to trace the further evolution of the void we
had to focus on the central regions of the void.  (In reality, as density
increases, a gradient of pressure would appear, which cannot be described
in the \lt model.)  Since superclusters are observed on the edges of the
voids, we interpret this singularity as an indicator of the presence of a
supercluster.

 \end{itemize}

Thus our numerical experiments were not entirely successful at generating a void
consistent with observational data, out of perturbations at last scattering that
would be as small as indicated by current structure formation theories.  Either
the initial perturbations had to be larger than the conventional values, $\delta
\rho / \rho \approx 10^{-5} \approx \delta V/c$, or else the present-day void
was too shallow.  This discrepancy calls for an explanation.  In attempts to
explain it, several hypotheses might be considered.  We list a set of such
hypotheses in the order of decreasing probability of being correct; this is of
course our subjective evaluation.

 \begin{enumerate}

 \item   The experiments, here and in Paper II (Krasi\'nski and Hellaby
2004), showed that the final state is sensitive not just to the
amplitude of the initial velocity perturbation, but also to its profile
(the density perturbation is less significant).  We may not have identified
the profile that gives the best consistency with observations.  
 Work on this will be continued. We tried one approach already, but the
results were not encouraging. We took the density distribution in the void
at present as given, and the density distribution at recombination
homogeneous. We calculated the implied velocity distribution at
recombination, which was 40 times too large (see Sec.
\ref{modelfit}). Then we numerically decreased the values of that velocity
by the factor of 40, took it as part of the input data, and calculated the
implied density distribution at recombination. It was again too large, so we
decreased it numerically by the appropriate factor, took it as input, and
calculated the implied velocity field at recombination. This iteration
quickly converged to stable values of the amplitudes: $6 \cdot 10^{-5}$ for
density and $ 7 \cdot 10^{-3}$ for velocity. Clearly, this does not solve
our main problem.

 \item   Voids may not be as empty as they appear.  It is possible they
contain a significant amount of unobserved matter, such as gas or other baryonic
``dark matter". A present day density contrast of smaller absolute value is
easier to produce with small initial fluctuations. 

 \item   Matter may have more components than comoving dust, and these other
components may be dynamically significant.  Including other components into a
fully non-linear description, and allowing for the possibility that the various
kinds of matter do not comove, requires finding solutions of Einstein's
equations with two or more independent streams of matter as a source, which is
an extremely challenging problem.  Some possible matter components are:
         \begin{enumerate}
         \item   Cold dark matter, i.e. some form of non-baryonic matter that
decouples from normal matter very early on and starts forming structures before
last scattering.  The amplitude of such fluctuations could be larger than what
CMB measurements allow for baryonic matter.  Whilst there is evidence for some
non-luminous matter from galaxy rotation curves, galactic interactions and
gravitational lensing, it is less than the required ``concordance" value, and of
unknown composition.  The difficulty with this hypothesis is that cold dark
matter is not based on any confirmed physical theory and has yet to be detected.
         \item   A radiation component that is still significant just
after last scattering.
         \end{enumerate}

 \item   Observations give the Galaxy distribution in redshift space, and void
sizes are deduced by using a Hubble law based on a homogeneous Friedmann model.
Since the density distribution thus obtained is clearly not homogeneous, the use
of a Friedmann model is really not correct.  This may introduce quite significant
errors in the density and velocity profiles of voids.  The well known ``finger
of God" effect is an example of a large discrepancy between true positions, and
the Friedmann-based mapping from redshift space to physical distances.
Consistency with Einstein's equations requires use of an inhomogeneous
cosmological model to correctly map redshift space into physical distances and
thus determine how velocities vary with distance. At present, neither the
density distribution nor the velocity field of galaxies is known with
confidence.\footnote{The velocities are measured, but then, without the
inconsistent assumption of homogeneity, we do not know to which points in the
curved manifold of the Universe they should be attached.} Had we known both
these fields with a reasonable precision, we might use the L--T model to {\it
calculate} the fluctuations in density and velocity at recombination.

 \item   Reliable measurements of temperature fluctuations of the cosmic
microwave background (CMB) radiation are available only for angular
scales larger than 0.5$^{\circ}$ (see Fig, \ref{wmap}), with
the scatter becoming large at 0.2$^\circ$.  On the other
hand, as shown in Paper I (Krasi\'nski and Hellaby 2004) and in sec.
\ref{ladiam}, the angular diameters, in the CMB sky, of structures the
right size to evolve into voids, are less than 0.25$^{\circ}$.
Though current theory suggests the amplitude decreases at smaller scales,
this has yet to be confirmed by improved CMB temperature measurements.

 \item   Solid observational data give the temperature fluctuation of
the cosmic microwave background radiation at present, $\Delta T/T$, separately
for different modes of perturbation.  However, there are several factors
contributing to $\Delta T/T$ (see section \ref{infl}), which may partially
cancel each other.  We were interested in the magnitude of the fluid velocity at
last scattering, as an initial condition for our models, but we were not able to
locate a formula relating the fluid velocity to the observed $\Delta T/T$.  Thus
our estimate of $\Delta V$ may be inaccurate.  We ourselves do not plan to enter
this field, however we would welcome an explicit calculation with a reliable
result.

 \item   The Universe may be much older than currently believed, making the time
available for void formation much longer.  This would require adjustment of the
matter content of the universe, of the value of the cosmological constant or of
the value of the Hubble constant, each of which may affect the
structure-formation timescale. 

 \item   General Relativity (GR) may not be the right theory of the evolution
of the Universe.  There is certainly no lack of proposed modifications or
alternatives.  While some would gladly accept this conclusion, and no-one can
claim with certainty that GR will survive all future scrutiny, we wish to stress
that GR has a much stronger experimental basis than any cosmic structure
formation theory.

 \end{enumerate}

\section*{ACKNOWLEDGMENTS}

KB wishes to thank the following people who helped him a lot in the work
on his MS Thesis, on which this paper is based: his parents for their
love and support, his supervisor Andrzej Krasi\'nski for his help with
understanding the subject, great patience and heartily smile,
Paulina Woj\-cie\-chow\-ska for all the helpful suggestions and
comments, Charles Hellaby, Jacek Jezierski, Fiona Hoyle, Krzysztof Jahn, 
Mi\-cha\-\l{} Cho\-do\-ro\-wski, Ewa \L{}okas and 
Roman Jusz\-kie\-wi\-cz.
 CH thanks the South African National Research Foundation for a Core grant,
and for funding from the Poland-South Africa Technical Cooperation Agreement.
 AK wishes to thank C. Hellaby and the Department of Mathematics and
Applied Mathematics of the Cape Town University, where this paper was
finalised, for hospitality, and the NRF of South Africa for financing
his visit out of the Agreement mentioned above.

\bsp

\label{lastpage}

\end{document}